\documentclass[showpacs,twocolumn,aps,floatfix,superscriptaddress,noshowpacs]{revtex4}
\usepackage[mathlines]{lineno}

\usepackage{amsthm,amsmath,amssymb,graphicx,bm,color,mathrsfs,verbatim,epstopdf,dcolumn,dsfont,cancel}

\usepackage{algorithm}
\usepackage[noend]{algpseudocode}
\makeatletter
\def\BState{\State\hskip-\ALG@thistlm}
\makeatother

\usepackage{graphicx}

\graphicspath{{figs/},{./}}

\usepackage{hyperref}
\hypersetup{ 
	colorlinks = true
}

\usepackage{color}

\begin{document}


\title{Reinforcement learning for autonomous preparation of Floquet-engineered states: \\ Inverting the quantum Kapitza oscillator}%

\author{Marin Bukov}
\email{mgbukov@berkeley.edu}
\affiliation{Department of Physics, University of California, Berkeley, CA 94720, USA}

\begin{abstract}
I demonstrate the potential of reinforcement learning (RL) to prepare quantum states of strongly periodically driven non-linear single-particle models. The ability of Q-Learning to control systems far away from equilibrium is exhibited by steering the quantum Kapitza oscillator to the Floquet-engineered stable inverted position in the presence of a strong periodic drive within several shaking cycles. The study reveals the potential of the intra-period (micromotion) dynamics, often neglected in Floquet engineering, to take advantage over pure stroboscopic control at moderate drive frequencies. Without any knowledge about the underlying physical system, the algorithm is capable of learning solely from tried protocols and directly from simulated noisy quantum measurement data, and is stable to noise in the initial state, and sources of random failure events in the control sequence. Model-free RL can provide new insights into automating experimental setups for out-of-equilibrium systems undergoing complex dynamics, with potential applications in quantum information, quantum optics, ultracold atoms, trapped ions, and condensed matter.

\end{abstract} 

\date{\today}
\maketitle


\section{Introduction}

\begin{figure}[t!]
	\includegraphics[width=1.0\columnwidth]{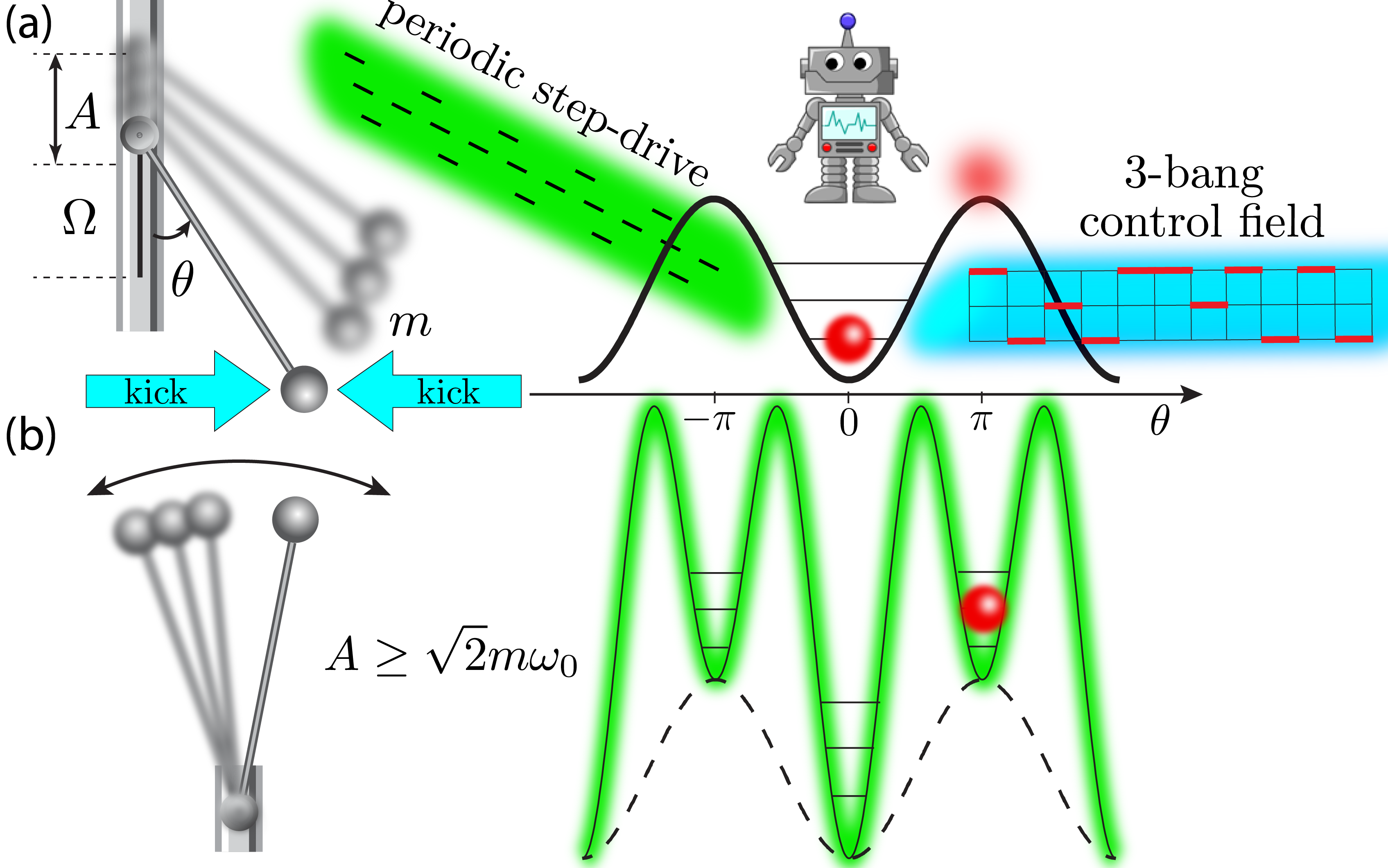}
	\caption{\label{fig:kapitza} Prototypical example of Floquet engineering: high-frequency periodic modulation stabilizes the metastable equilibrium at the inverted position of a classical pendulum (left) and a quantum oscillator (right). (a) schematic Floquet control setup displaying the step-periodic drive (green), and the control protocol (cyan). The purpose of this work is to prepare states localized at the inverted position (b) in the presence of strong periodic modulation without knowledge about the physical system, using Reinforcement Learning [\href{https://mgbukov.github.io/movies/RL_kapitza/movie-1.mp4}{Video 1}].}
\end{figure}

The use of strong periodic modulations to design properties of quantum matter is an established approach from the quantum simulation toolbox. Commonly known as \emph{Floquet engineering}~\cite{goldman2014periodically,bukov2015universal,eckardt2017atomic}, these ideas prove essential to realize states of matter inaccessible in conventional materials. Prominent achievements include stabilizing unstable equilibria~\cite{kapitza_51,lerose2018quantum,citro2015dynamical} [Fig.~\ref{fig:kapitza}], dynamical localization and the related dynamically-controlled Mott insulator-superfluid transition in ultracold optical lattices~\cite{eckardt_05,sias_08,zenesini_10}, the emulation of strong artificial gauge fields~\cite{jaksch_03,struck_11,struck_12,hauke_12,struck_13,aidelsburger_13,miyake_13,atala_14,goldman_gaugefields_14,kennedy_15,roushan2017chiral}, imprinting topological and spin properties into band insulators~\cite{oka_09,kitagawa_11,rudner_13,jotzu_14,aidelsburger_14,flaeschner_15,tarnowski2017characterizing,aidelsburger2017artificial,jotzu_15}, topological defects~\cite{tarnowski_17}, quantum magnetism~\cite{eckardt_10,mentink2015ultrafast,meyer_17,gorg2017enhancement}, spin-orbit coupling~\cite{anderson_13,galitski_13,jimenez-garcia_15}, synthetic dimensions~\cite{celi_13,stuhl2015visualizing,mancini2015observation}, and photonic topological insulators~\cite{rechtsman_13,hafezi_14,mittal_14}.

The current bottleneck in Floquet engineering is caused by detrimental heating effects, due to uncontrolled energy absorption as a result of a proliferation of Floquet many-body resonances, beyond the inverse-frequency expansion~\cite{bukov_15_res,claeys2017spin,arze2018out}. The short-time dynamics of weakly-interacting bosons was shown to be dominated by parametric resonances~\cite{bukov_15_prl,lellouch_17,lellouch_18,naeger2018parametric,boulier2018parametric}. Fermi's Golden rule was applied to the long-time evolution, and leads to a featureless infinite temperature state~\cite{bilitewski_15,bilitewski_15b,reitter_17}. Theoretically, for non-integrable many-body lattice systems with bounded on-site Hilbert space dimension, heating was proven to be (at least) exponentially suppressed in the drive frequency~\cite{abanin_17,mori_15,kuwahara2016floquet}, which allows for the formation of transient long-lived prethermal steady states, ideally suited for Floquet engineering~\cite{weidinger2017floquet,peronaci2017resonant,howell2018frequency,messer2018floquet}. 

Knowing how to Floquet-engineer an exponentially long-lived state of matter leaves the important open problem of how to steer the system in the desired target state. The state-of-the-art approach to manipulate periodically-driven systems is the adiabatic variation of parameters~\cite{guerin2003control,weinberg2017adiabatic,novicenko_17,ho2016quasi}. While drive-induced photon absorption avoided crossings in the quasienergy spectrum need to be passed quickly in order to avoid spending much time on resonance and heating up, the state should go slowly, compared to the inverse energy gap, through conventional avoided crossings to suppress excitations~\cite{weinberg2017adiabatic}. This tension leads to the breakdown of Floquet adiabatic perturbation theory~\cite{weinberg2017adiabatic,hone_97,eckardt_08}, despite the existence of parametrically-controlled windows of applicability, and points towards the necessity to develop new approaches for Floquet control in both single-particle and many-body systems.

Reinforcement Learning (RL)~\cite{sutton1998reinforcement} is one of the most promising techniques in Machine Learning~\cite{bishop2006pattern,dunjko_17,ML_review}, closely related to optimal~\cite{rabitz_00,todorov2006optimal,glaser_15,grape_05,caneva_11,sorensen2018quantum,peruzzo2014variational,mcclean2016theory}, feedback~\cite{doherty_99,gillett_10,wiseman_94,rabitz2009focus,frank2017autonomous} control, and evolutionary algorithms used in quantum chemistry and optics to learn molecular control~\cite{judson1992teaching,baumert1997femtosecond,yelin1997adaptive,pearson2001coherent,weiner2000femtosecond,bartels2000shaped,meshulach1998coherent,herek2002quantum,weinacht1999toward,bardeen1997feedback,levis2001selective,assion1998control,nuernberger2010femtosecond,brixner2003quantum,winterfeldt2008optimal}. It is especially well-suited to \emph{autonomously} control systems in the presence of strong drives since it is model-free and robust to imperfections and noise. In physics, RL has been used to navigate thermals~\cite{reddy2016learning} and turbulent flows~\cite{colabrese_17}, design experiment setups in quantum optics~\cite{melnikov2018active}, construct molecules with prescribed properties~\cite{popova2017deep}, and to control quantum systems~\cite{dong2008incoherent,chen_14_ML,bukov_17RL,august_18,foesel_18,zhang2018automatic,niu2018universal,albarran2018measurement}. Without a physical model, RL was shown to produce comparable results to algorithms from optimal control~\cite{bukov_17RL}. Ideas from quantum physics have been suggested to improve RL-related algorithms~\cite{paparo2014quantum,sriarunothai2017speeding,clausen2016quantum}. Despite recent progress, RL's potential remains massively unexplored in physics. 

Inspired by Ref.~\cite{bukov_17RL}, the present work demonstrates the suitability of RL to study the nonequilibrium quantum dynamics of strongly-driven Floquet-engineered states. In a numerical simulation of a quantum experiment, starting with zero knowledge about the system, the RL agent learns how to optimally prepare inverted position states in the Kapitza pendulum from tried protocol configurations [see \href{https://mgbukov.github.io/movies/RL_kapitza/movie-1.mp4}{Video 1}]. The algorithm is applied to both the quantum and the classical oscillator. Unlike Ref.~\cite{bukov_17RL}, the agent learns from quantum (i.e.~non-deterministic) measurement data, and is shown to remain robust after adding noise to the initial state, and occasional random failure events in the control sequence. The study shows the advantage of exploiting the micromotion dynamics for control to achieve higher fidelities compared to stroboscopic control.

\section{Floquet Control Problem}

Consider the Hamiltonian of the horizontally kicked quantum Kapitza oscillator
\begin{eqnarray}
H(t) &=& H_0 + H_\mathrm{drive}(t) + H_\mathrm{control}(t),\nonumber\\
H_0 &=& \frac{p_\theta^2}{2m} - m\omega_0^2\cos\theta ,\nonumber\\
H_\mathrm{drive}(t) &=& - \frac{A}{2m}\mathrm{sign}(\cos\Omega t)\left( \sin\theta\; p_\theta + p_\theta \sin\theta \right) \nonumber\\
&& - \frac{A^2}{8m}\left(1-\mathrm{sign}(\sin2\Omega t)\right)\cos 2\theta,\nonumber\\
H_\mathrm{control}(t) &=& h(t)\sin\theta, 
\label{eq:kapitza_H}
\end{eqnarray}
where $m$ and $\omega_0$ are the mass and natural frequency of the oscillator $H_0$ with position (angle) and (angular) momentum variables obeying $[p_\theta,\theta]\!=\!-i$. Applying a strong vertical periodic drive of constant amplitude $A$ and frequency $\Omega\!=\!2\pi/T$ is known to stabilize the metastable inverted position at $\theta\!=\!\pi$, a paradigmatic example of Floquet engineering~\cite{bukov2015universal} [Fig.~\ref{fig:kapitza}]. The latter requires the Floquet drive to couple strongly to the system, with an amplitude scaling linearly with $\Omega$ in the lab frame~\cite{bukov2015universal}. To see how stabilization occurs, and to avoid working at large amplitudes, it is advantageous to adopt the rotating frame description~\eqref{eq:kapitza_H}, at the expense of introducing a second harmonic in $H_\mathrm{drive}(t)$ [cf.~App.~\ref{app:kapitza_theory}]. Therefore, the piece-wise constant drive, designed to speed up numerical simulations, repeats every four steps. The units are chosen such that $p_\theta$, $\theta$, $m\omega_0$, $A$ are all dimensionless, and $\hbar=1$. Energy is measured in units of $\omega_0$.

For $h(t)\equiv 0$, the dynamics of the uncontrolled Kapitza oscillator at integer multiples of the drive period $T$ (i.e.~stroboscopically) is exactly described by the Floquet Hamiltonian $H_F(\Omega)$. In the infinite-frequency limit, taking the period-average of Eq.~\eqref{eq:kapitza_H} gives
\begin{equation}
H_F(\Omega\to\infty,h\equiv 0)=\frac{p_\theta^2}{2m} - m\omega_0^2\cos\theta - \frac{A^2}{8m}\cos 2\theta.
\label{eq:H_F}
\end{equation}
The periodic drive renormalizes the potential energy of the oscillator [Fig.~\ref{fig:kapitza}b] and, whenever $A\!\geq\!\sqrt{2}m\omega_0$, the potential supports a stable equilibrium at $\theta\!=\!\pi$ with frequency of harmonic oscillation $\omega'\!=\!\sqrt{A^2/(2m^2) \!-\! \omega_0^2}$. Away from the limit $\Omega\!\to\!\infty$, finite-frequency corrections can be incorporated using the inverse-frequency expansion~\cite{bukov2015universal}, yet the exact form of $H_F(\Omega)$ remains unknown. This leads to a modification of the critical amplitude, yet the stabilizing behavior persists qualitatively  down to $\Omega\!\gtrsim\! 6\omega_0$ for the step-drive.  

Turning on the control field $h(t)$ compromises the time-periodicity of $H(t)$, and one can in general no longer rely on Floquet theory. The unknown control field $h(t)\in\{0,\pm 4\}$, $t\!\in\![0,t_f]$, of duration $t_f\!=\!N_T T$ is built from a sequence of constant horizontal momentum kicks of duration $\delta t$ exerted on the oscillator, called bangs, to speed up the efficiency of the RL algorithm. The bounded kick strength reflects possible constraints in experiments with too large control fields; the exact values $\{\pm 4\}$ are chosen to be on the same order of magnitude as the bare oscillator frequency, so that no term dominates the Hamiltonian and the dynamics cannot be studied using perturbation theory. The $0$-bang allows to turn off the control field. I further consider drive-commensurate protocols of two types: (i) stroboscopic: $T\!=\!\delta t$, and (ii) commensurate non-stroboscopic, $T\!=\!4n\delta t$, $n\in\mathbb{N}$ [Fig.~\ref{fig:kapitza}]. To control the Kapitza oscillator, non-stroboscopic protocols are chosen $N_T\!=\!15$-driving-cycles long, with $8$ steps per cycle ($120$ bangs). The protocol space contains $3^{120}\!\sim\!10^{57}$ configurations.

The objective of this study is to determine a bang-bang protocol $h(t)$ which finds the system in the ground state $\vert\psi_i\rangle$ of the non-driven uncontrolled Hamiltonian $H_0$, and brings it as close as possible to the target state $\vert\psi_\ast\rangle$ -- the eigenstate of the finite-frequency $H_F(\Omega)$ localized at the inverted position, in a fixed amount of time. The figure of merit is the fidelity$F_h(t_f)\!=\!\vert\langle\psi(t_f)|\psi_\ast\rangle\vert^2$ of being in the target state at the end of the control sequence. Note that the amplitude $A$ of the instantly turned on periodic drive is held constant during control and, therefore, there is no natural adiabatic path [in $h$-space] between the initial and target states. There is no small parameter to do perturbation theory in, either.

\begin{figure}[t!]
	\includegraphics[width=1.0\columnwidth]{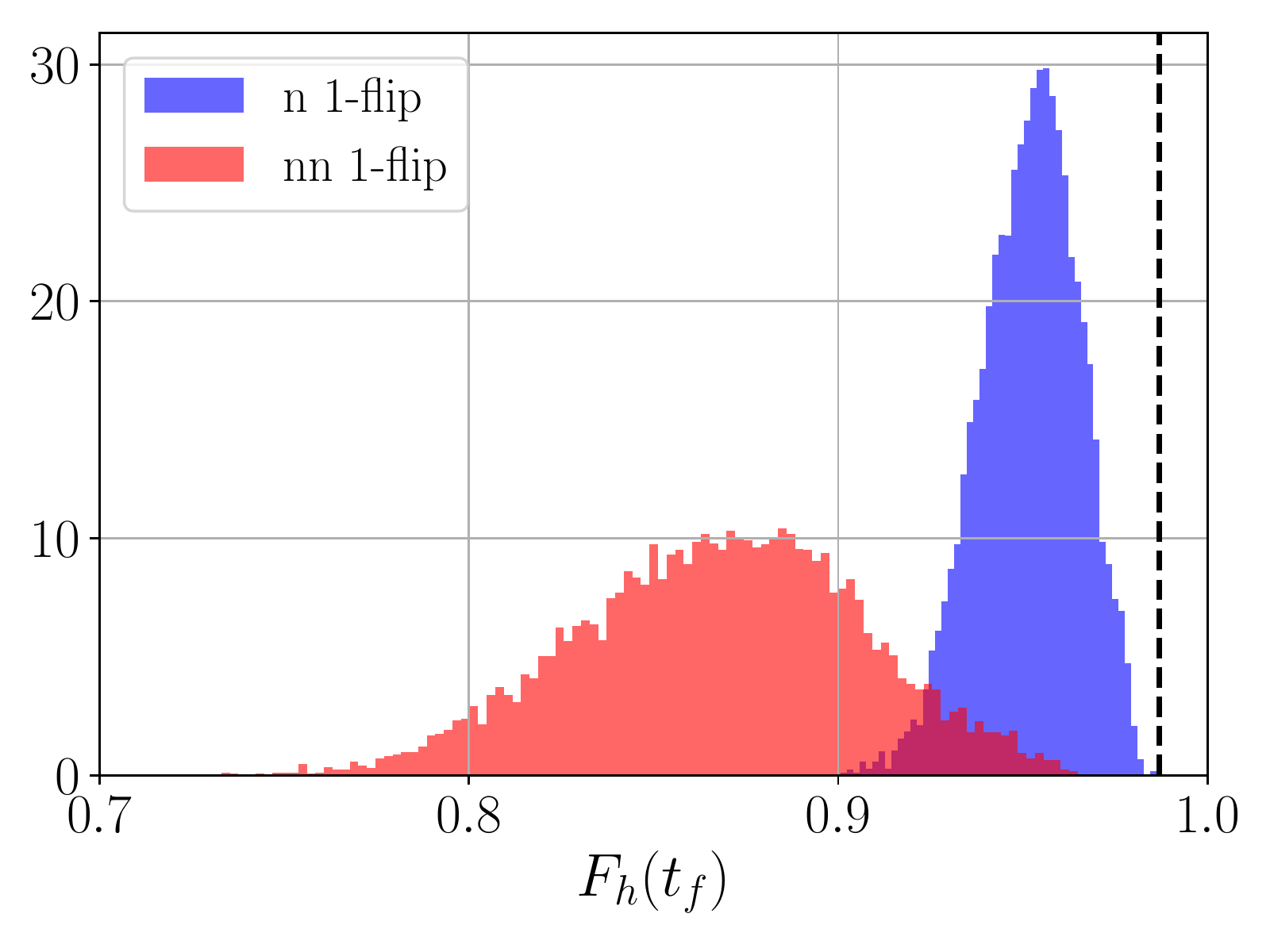}
	\caption{\label{fig:SD_excitations}Distributions of the nearest (n) [i.e.~$\{-4,0\}$ and $\{0,+4\}$] and next-nearest (nn) [$\{-4,+4\}$] $1$-flip excitations of the SD protocols [see text]. The dashed vertical line marks the fidelity of the absolute maximum. The data refers to the sample of all local protocol minima which satisfy $F_h(t_f)>98\%$]. The model parameters are the same as in Fig.~\ref{fig:reward_vs_episode}.}
\end{figure}

\section{Complexity of the Deterministic Kapitza Control Problem}

At present, I am not aware of an analytical theory to find the optimal protocol or even predict its fidelity in the Kapitza control problem. It is, therefore, advantageous to acquire some intuition about the properties of the control landscape~\cite{rabitz2004quantum}, which has recently been shown to exhibit a variety of phase transitions~\cite{bukov_17RL,solon_17,larocca2018quantum} including glassy control phases~\cite{day2018glassy} and symmetry breaking~\cite{bukov2017broken} in non-Floquet control problems. 

To begin with, a quick check shows that using random protocol sequences leads to about $10\%$ fidelity [App.~\ref{app:protocols}] -- this is expected to reflect the performance of the RL agent during the first episodes of training, when it is still unfamiliar with the behavior of the physical system. 

To get an estimate of the magnitude of the reachable fidelities for the chosen set of model parameters, I use Stochastic Descent (SD), initiated from random protocol configurations, and flip the protocol bangs randomly one at a time, until a local minimum of the infidelity landscape is reached~\cite{day2018glassy,bukov_17RL}. Such minima represent locally optimal protocols, close to which greedy optimization algorithms are likely to get stuck, due to the glassy character of control landscapes~\cite{day2018glassy}. The obtained sample of $10^6$ SD-protocols has mean fidelity $87\%$ [App.~\ref{app:protocols}]. Out of these, ninety-two local minima protocols have fidelity greater than $98\%$, with the absolute maximum at $98.68\%$. The bang sequences of the corresponding protocols are, however, completely different which suggests that they may occupy deep pockets in the control landscape (w.r.t.~the Hamming distance)~\cite{rabitz2004quantum}. 

To test this assertion, starting from each one of the ninety-two best SD minima, I compute the fidelities of all 1-flip variations [called excitations], which fall in two categories: nearest (n) flips are those between protocol values $\{-4,0\}$ and $\{0,+4\}$, while next-nearest (nn): $\{-4,+4\}$. Figure~\ref{fig:SD_excitations} shows the distributions of one-flip excitations. Intuitively, one expects that the wavefunction cannot undergo drastic changes within the short kicks of strength ($|h|\delta t\!=\!0.314$) during a single bang. Nonetheless, on average the fidelity of $1$-flip excitations drops by more than $10\%$, which points at the rugged profile of the Floquet control landscape. The total number of fidelity evaluations required to obtain the local SD-minima sample is about $10^8$ [$10^6$ runs, with on average $10^2$ evaluations each]. SD is a deterministic algorithm [i.e.~relies on the exact fidelities to operate]. In this work, I present an \emph{autonomous} RL algorithm which can be coupled to realistic experiments with multiple generic sources of noise.
 
\section{Challenges in Autonomous Quantum Control}

All information about a quantum state is encoded in its wavefunction. Hence, when developing algorithms for realistic experimental setups, the main challenge for autonomous control arises from the lack of access to the quantum states which are \emph{unmeasurable} mathematical constructs. Invoking Picard-Lindel\"of's uniqueness theorem to find the evolution operator $U_h(t,0)$ which integrates Schr\"odinger's equation, given a fixed initial state $|\psi_i\rangle$, one can parametrize the accessible states at later times $|\psi(t)\rangle$ by the protocol sequence $h(t)$ up to time $t$:
\begin{eqnarray}
\label{eq:psi_h_correspondence}
\{ |\psi(t)\rangle\!&=&\!U_h(t,0)|\psi_i\rangle\! \}  \Leftrightarrow  \\
&& \{h(t')\!:\! H[h(t')], |\psi_i\rangle, t'\!\in\![0,t]\}\!=\!\mathcal{S}. \nonumber
\end{eqnarray}
Although this mapping is not one-to-one [a state may be accessible using different protocols], it offers a significant advantage: while quantum states cannot be measured, the applied protocols $h(t)$ can actually be kept track of. Hence, fixing the initial state, one can infer which state the system ought to be in at a later time, from the applied protocol. In fact, the minimal amount of information an autonomous algorithm can have about the controlled quantum system is the applied protocol sequences.

Another challenge in experimental quantum control is the nondeterministic character of projective quantum measurements. Since they destroy the state, one is allowed to measure only once, at the end of the protocol when the system has evolved into the state $\vert\psi(t_f)\rangle$. Projective measurements are modeled to return $1$ with probability set by $F_h(t_f)=\vert\langle\psi(t_f)\vert\psi_\ast\rangle\vert^2$ if the system is found in the target state $|\psi_\ast\rangle$, and $0$ otherwise [App.~\ref{app:quant_meas}]. Therefore, the algorithm seeks to maximize the fidelity $F_h(t_f)$, whose true value remains unknown at the time of measurement, and is only estimated from the data. During the optimization process, the situation is in fact much more complicated since, as the available protocol family is explored, the true probability $F_h(t_f)$ to determine the measurement, changes from one protocol to the next, as different protocols in general lead to different final states. 

Further challenges arise from (i) the inability to prepare the initial state $|\psi_i\rangle$ with certainty. Experiments only prepare the desired initial state within a given fidelity window. Additionally, (ii) in experiments one has to account for the occasional failure of the control apparatus: even though the algorithm may have requested the protocol sequence $h(t)$, the physical system might experience a slightly modified protocol $h'(t)$ instead. Depending on the sensitivity of the optimal solution, this could lead to drastic changes in the reachable fidelities~\cite{bukov_17RL,day2018glassy} [see Fig.~\ref{fig:SD_excitations}]. Every experiment, (iii), comes with its own imperfections and difficulties: the Hamiltonian $H(t)$, assumed to model the system, is often merely a simplified approximation, which compromises the unconditioned applicability of idealized simulations. Finally, even if all of the above were not present, (iv) there are various constraints imposed by the dynamics of the physical system of interest. In this respect, controlling Floquet systems remains an unsolved challenging problem of nonequilibrium dynamics. It is, thus, desirable to have a versatile algorithm which is capable of dealing with the above scenarios in an efficient way.

\section{Reinforcement Learning}

In this work, I adopt a modification of Watkins Q-Learning algorithm~\cite{watkins1992qlearning} to study the Kapitza control problem. An RL agent seeks to find an optimal protocol $h(t)$ to prepare the target state $|\psi_\ast\rangle$, \emph{without knowledge} about the controlled system~\footnote{Note that the agent is presented with the target state, and does not discover it.}. To do this, it episodically gains experience and uses it by taking actions to construct bang-bang protocols one bang at a time. These protocols are applied to a simulated quantum system (the environment), returning a reward to the agent: the estimated fidelity $F_h(t_f)$, computed based on the measurement record obtained so far. As the number of training episodes increases, the algorithm progressively correlates protocols with their fidelity, a process referred to as \emph{learning}. Even though in a simulation it is possible to learn from the exact fidelities, in order to better simulate a realistic quantum experiment, for each protocol $h(t)$, the algorithm stores two integers, corresponding to the number $m$ of protocol encounters, and the number $n$ of $1$'s in the output of the binary quantum measurement. From them, it computes the reward $r$ estimating the mean current fidelity $r\!=\!n/m$ of being in the target state (for a given protocol). The error estimate $E$ to be within the $2\sigma$ window, controls the number of repetitions used to gather measurement statistics. Hence, during the learning process, the agent has to deal with noisy rewards. These intrinsically quantum features obscure the learning process significantly close to the best attainable fidelities, where it is known that differences in the higher decimal places of the fidelity can play a decisive role~\cite{day2018glassy}.  

The information the agent has about the controlled system is encoded in the RL state space $\mathcal{S}$, which
I define using the correspondence~\eqref{eq:psi_h_correspondence}. For instance, for a $4$-step-long protocol $\{+4\}$,$\{+4,-4\}$,$\{+4,-4,0\}$,$\{+4,-4,0,0\}\!\in\!\mathcal{S}$ are all admissible RL states. Note that the size of $\mathcal{S}$ scales exponentially with the number of bangs. While this aggravates the exploration of the state space, it is well within the scope of RL algorithms to learn nearly-optimal policies in complex environments with exponentially-large state spaces, as becomes clear from recent success in mastering video and board games beyond human level~\cite{mnih2015human,silver2016mastering}. Importantly, this scaling does \emph{not} depend on the Hilbert space dimension of the quantum system which makes the algorithm applicable to large systems [one does need an experiment to simulate their dynamics to provide rewards, though]. In this respect, the RL algorithm is modular: the learning part [which can be chosen insensitive to the Hilbert space dimension] is separate from the quantum mechanics part [which provides the rewards and in a simulation would suffer from the limitations due to large Hilbert space dimensions]. To construct protocols on-the-fly, every time step the agent invokes its knowledge gathered so far to pick an action from the set $\mathcal{A}\!=\!\{-4,0,+4\}$ based on the predicted expected reward. I use an $\varepsilon$-greedy exploration policy, which is attenuated exponentially with the number of episodes~\cite{sutton1998reinforcement}. Finally, the reward space is $\mathcal{R}\!=\!\{ r\!\in\![0,1]\!:\! r\!=\!n/m\}$. The algorithm also applies experience replays to enforce exploration around the estimated best-encountered protocol, cf.~App.~\ref{app:algo}.
 
\begin{figure}[t!]
	\includegraphics[width=1.0\columnwidth]{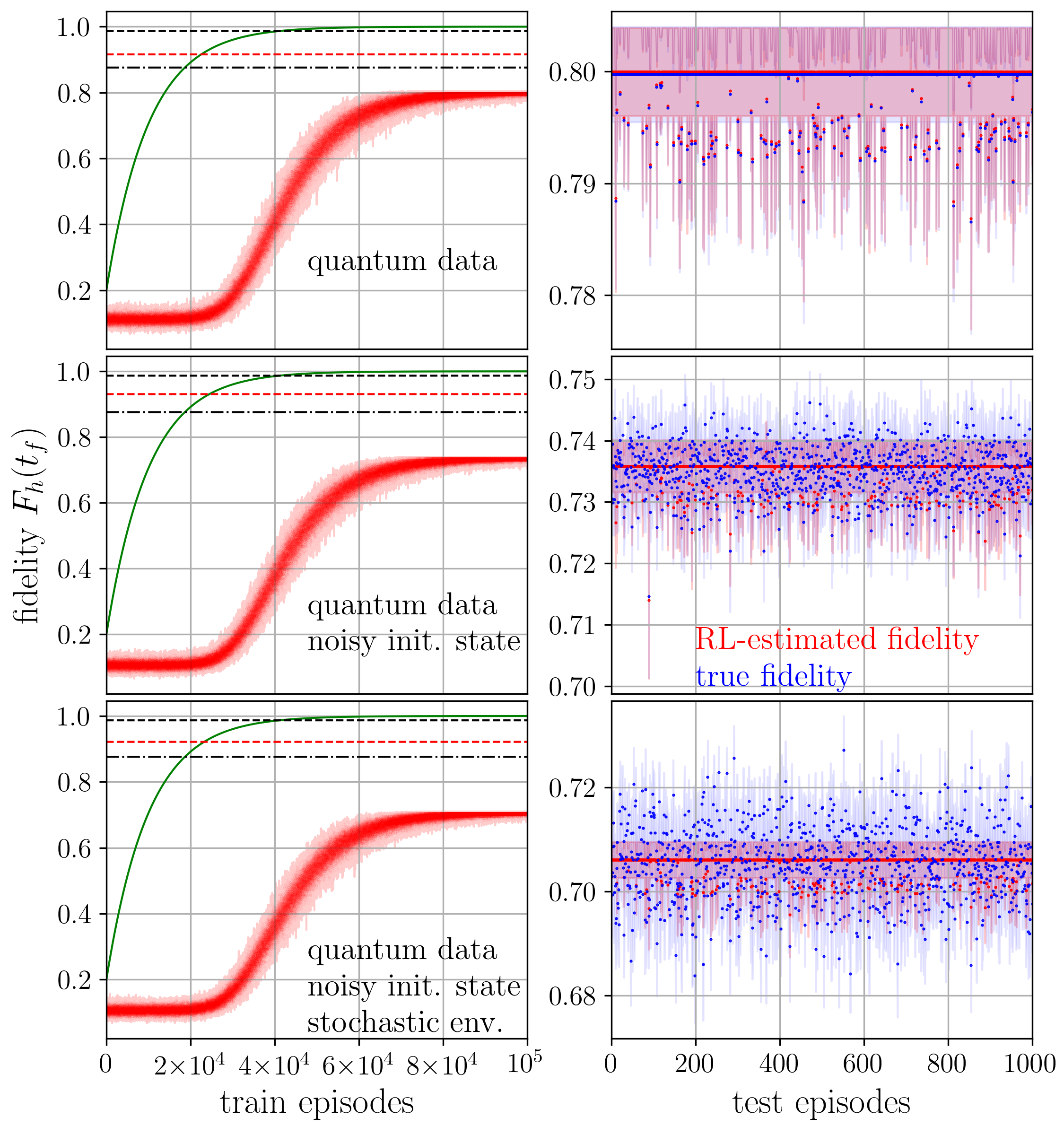}
	\caption{\label{fig:reward_vs_episode} Running fidelity estimate during Train (left) and Test (right) stages of Q-Learning for the quantum Kapitza oscillator. The three rows show data for stochastic rewards only (top), stochastic rewards and noisy initial condition (middle), and stochastic rewards, noisy initial condition and stochastic environment (bottom). The horizontal black lines show the average fidelity of local infidelity minima computed using one-flip SD (dashed dotted), and the best SD realization (dashed) out of $10^6$ local minima samples obtained  with \emph{deterministic cost function}. The horizontal red dashed line shows the best-encountered protocol using RL during the Train stage. The RL data points (red) are averaged over $100$ seed realizations of the pseudo-random number generator, with the uncertainty window (shaded area) computed using a bootstrapping approach. The green curve shows the exponentially attenuated exploration schedule $\varepsilon(n_\mathrm{ep})$, normalized within $[0,1]$ for display purposes: unity corresponds to no exploration. The oscillator parameters are $N_T=15$ periods with $8$ steps each, $\Omega/\omega_0=10$, $A=2$ and $m\omega_0=1$.}
\end{figure}

\section{Autonomously Inverting the Quantum Kapitza Oscillator}

The Q-Learning agent is first trained for $10^5$ exploration episodes, followed by $10^3$ greedy test episodes to examine the stability of the learning process. Due to the probabilistic character of $\varepsilon$-greedy exploration, the algorithm is run for $100$ distinct seeds of a pseudo-random number generator, and the results I present show averages, cf.~App.~\ref{subapp:single_vs_average}. Figure~\ref{fig:reward_vs_episode}a shows that the fidelity $F_h(t_f)$ of being in the target state increases consistently and then saturates at about $80\%$. Hence, the RL agent is capable of autonomously controlling the Kapitza oscillator by using only information from noisy quantum measurements. Note that the agent slightly overestimates the true fidelity (blue). Yet, it learns the correct noise correlations in the reward [test stage: blue and red shaded areas].

To demonstrate the robustness of RL to noise in the initial state, I draw a Haar-random state $\vert\phi\rangle$, and consider the noisy initial state $\vert\psi_i(\eta)\rangle \!\propto\! \vert\psi_i\rangle \!+\! \eta\vert\phi\rangle$, with $\eta\!=\!0.31$ such that $\vert\langle\psi_i|\psi_i(\eta)\rangle\vert^2\!\approx\!0.9$. This results in a small drop of fidelity to about $73\%$, parametrically controlled by $\eta$ [Fig.~\ref{fig:reward_vs_episode}b]. Therefore, Q-Learning is stable to small perturbations in the initial condition. Additionally, I expose the RL agent to a stochastic environment, in which every bang is randomly replaced by any of the three available actions with probability $\zeta\!=\!1/120$, mimicking occasional spontaneous failure in the control apparatus. The value $\zeta\!=\!1/120$ is chosen so that one bang of the $120$-bang-long control sequence fails on average. This scenario, similar to gate failure in quantum computing, expectedly leads to a further reduction to about $71\%$ [Fig.~\ref{fig:reward_vs_episode}c]. Hence, RL is also capable of learning in stochastic quantum environments. Interestingly, the agent is capable of de-noising the experimental rewards, as indicated by the uniform envelope of the estimated (red) compared to the true (blue) fidelity fluctuations.

To investigate the ability of the algorithm to navigate noisy environments, I also trained the agent in a noiseless deterministic environment, but tested it in a noisy stochastic setup, cf.~App.~\ref{subapp:noise_exp}. The test-stage true fidelity fluctuates about the same value, as if the agent was trained in a noisy stochastic environment. This behavior likely originates from the intrinsic exploration noise built in $\varepsilon$-greedy policy. The inability to exploit the knowledge in the presence of initial state noise is presumably a consequence of the RL state space definition~\eqref{eq:psi_h_correspondence}. This reveals a potential drawback: if the system is initiated in a sufficiently different state, the gained knowledge is not immediately exploitable, and the agent takes time to explore again [App.~\ref{subapp:noise_exp}]. This is, however, a feature of the current choice for the state-action space, rather than of the algorithm. 
 
I could not distinguish any significant features in the best protocol sequences [cf.~App.~\ref{app:protocols}], which suggests that the bang-bang family might not be naturally suitable for Floquet control problems. Nonetheless, one can visualize the dynamics of the real-space probability distribution of the oscillator. I consider three stages: (i) the Floquet system is subject to the best RL protocol in the presence of the Floquet drive. Once the control stage is over, (ii) I keep the Floquet-drive on with $h\!\equiv\!0$, before (iii) both the Floquet drive and the control are turned off ($h\!\equiv\!0$, $A\!=\!0$) and the system evolves under $H_0$, see \href{https://mgbukov.github.io/movies/RL_kapitza/movie-1.mp4}{Video 1} [best-encountered RL] and \href{https://mgbukov.github.io/movies/RL_kapitza/movie-2.mp4}{Video 2} [best SD 1-flip local minimum]. Once can observe the complexity of preparing entire local probability distributions with a single global control field, as becomes clear from the large fluctuations in fidelity between the short bangs. Note that the RL agent seems to first push the real-space weight of the wavefunction clockwise [\href{https://mgbukov.github.io/movies/RL_kapitza/movie-1.mp4}{Video 1}], before the final state is eventually reached from the opposite counter-clockwise direction. This is reminiscent of the classical problem with scarce control resources (mountain car paradigm in RL~\cite{sutton1998reinforcement}) where, in the short time available, one might decide to push the pendulum one way to convert energy from the drives into potential energy which, with the help of gravity, can then be unleashed to reach the inverted position from the other side. The quantum character of the dynamics likely determines the precise nontrivial bang sequence to keep the structure of the wavepacket during the evolution. This classical-like behavior is intriguing, since the quantum nature of the dynamics is clearly exhibited during stage (iii), where the overlap with the target state remains large even when the oscillator is not controlled or driven, as a consequence of $\vert\psi_\ast\rangle$ having a finite overlap with the excited eigenstate of $H_0$ corresponding to the meta-stable classical inverted state. 

\begin{figure}[t!]
	\includegraphics[width=1.0\columnwidth]{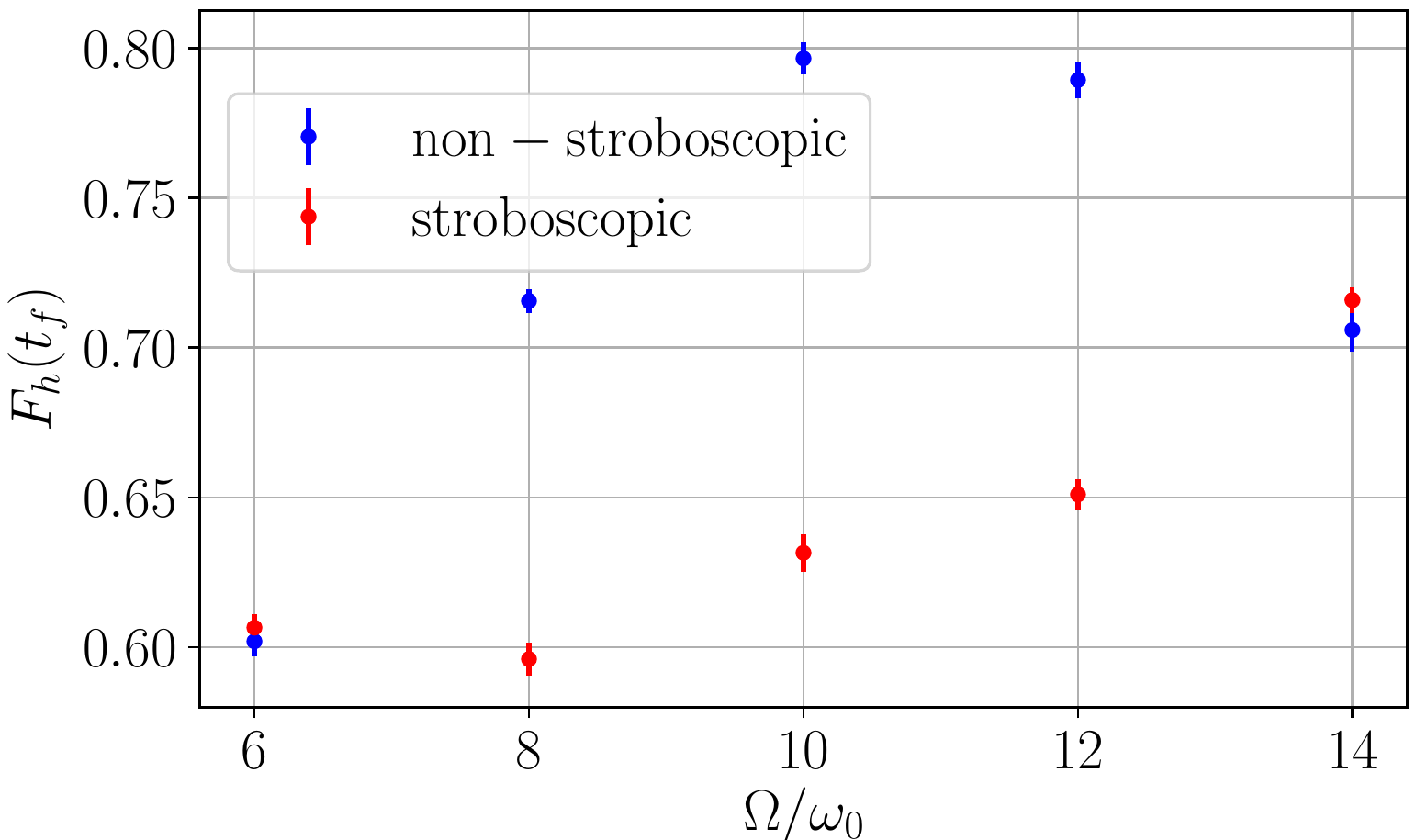}
	\caption{\label{fig:strobo_vs_constrobo}RL agent performs better in non-stroboscopic ($8$ bangs per cycle) than stroboscopic ($1$ bang per cycle) control at moderate drive frequencies. See Fig.~\ref{fig:reward_vs_episode} for the parameters.}
\end{figure}

Last, the study also confirms the clear superiority of non-stroboscopic Floquet control ($8$ bangs per cycle) over stroboscopic control ($1$ bang per cycle). Figure~\ref{fig:strobo_vs_constrobo} shows the learned saturation fidelity for several moderate frequencies in both cases. As expected, since the stroboscopic protocols can also be viewed as non-stroboscopic ones, the stroboscopic optimal fidelity is always a lower bound on the non-stroboscopic one. However, the smaller stroboscopic family ($3^{15}\!\sim\! 10^7$ protocols) can be explored more efficiently which leads to seemingly better RL performance at high drive frequencies. Even though nonstrobscopic dynamics is often neglected in Floquet engineering, it can offer advantages in Floquet control, and should not be easily dismissed~\cite{bukov_14_pra,desbuquois_17}.

\section{Discussion}

The best SD-protocol out of the family of local minima has $98.6\%$ fidelity [Fig.~\ref{fig:reward_vs_episode}, dashed black line], better than the average learned RL-fidelity. It is also superior to the best protocol encountered by the RL agent during training at $91.6\%$ [Fig.~\ref{fig:reward_vs_episode}, dashed red line]. Why could the agent not learn any of these protocols? One possibility is that it estimated their true fidelities poorly, and decided to ignore them. Another suggests the existence of very deep pockets in the infidelity landscape, which are unstable to noise, naturally present in the exploration schedule [Fig.~\ref{fig:SD_excitations}]. RL is designed to find stable solutions, even if they are further from the global minimum as measured by the cost function. Last but not least, the RL fidelities do improve with increasing the number of training episodes to $10^6$ -- still a tiny fraction of the total RL state space size, cf.~App.~\ref{subapp:episode_scaling}. In this respect, notice that the data in Fig.~\ref{fig:reward_vs_episode} is shown for $10^5$ fidelity evaluations, as opposed to $10^8$ evaluations for SD \footnote{Without counting non-exploratory repetitions required to collect measurement statistics}.

The control setup considered in this paper comes in contrast to typical Floquet control problems, which support an adiabatic path in parameter space between the initial and target states, e.g.~by slowly turning on the drive amplitude. 
As a result, comparing the numerically obtained protocols to analytical predictions is a formidable challenge. However, this challenge does not arise from the specific nonadiabatic setup alone. Even though adiabatic perturbation theory has been extended to periodically-driven systems, Floquet resonances, leading to gaps in the quasienergy spectrum along the adiabatic trajectory, are known to result in the breakdown of Floquet adiabaticity~\cite{hone_97,eckardt_08,weinberg2017adiabatic}. In contrast, this is not a problem in static (i.e.~non-Floquet) systems, where RL has been applied to a many-body spin chain to prepare ground states, adiabatically connected by the control field~\cite{bukov_17RL}: it was found that, at short durations, the functional form of optimal protocols differs significantly from that of adiabatic protocols, but can still be understood within the analytical framework of shortcuts to adiabaticity. Extending such ideas to Floquet systems is, to be best of my knowledge, an open problem, where RL and optimal control could provide useful insights.

One might also raise a valid objection that experiments currently cannot project the system to exact Floquet eigenstates, and hence using this particular target state makes the study not immediately applicable to realistic experimental setups. It is, however, possible to target a quasi-Gaussian state, localized at the inverted position, $\langle\theta|\psi_\ast\rangle\!\propto\!\exp(-\left(\omega'\right)^{1/4}\cos\theta)$. This gives qualitatively similar results, cf.~App.~\ref{app:gaussian_target}. 

It is also important to mention that there exist alternative algorithms that can be used to study Floquet control. Examples include GRAPE~\cite{grape_05}, CRAB~\cite{caneva_11} , VQE~\cite{peruzzo2014variational,mcclean2016theory,dimitrescu2018cloud}, QAOA~\cite{farhi2014quantum,ho2018ultrafast,ho2018efficient}, and Lyapunov-based feedback control~\cite{cong2013survey,grivopoulos2003lyapunov,zhao2009implicit,qamar2017lyapunov}. Whereas some of them require to have a model for the system under control, RL is model-free and can be applied in situations where the Hamiltonian of the system (more generally, the dynamics of the environment) is unknown. In the current study, I make use of a model only to provide the training data. 

Quite generally, bang-bang protocols also come with an experimental limitation, posed by the bandwidth of pulse generators. Even though they constitute a convenient theory starting point, it would be interesting to parametrize the protocols by Fourier components, an idea underlying the CRAB algorithm~\cite{caneva_11}. Such protocols can be resonant with the drive, and the corresponding control process likely admits a Floquet description. I should emphasize that the use of bang-bang protocols is \emph{not} at all a requirement imposed by RL algorithms, and certain RL algorithms (e.g.~Policy Gradient) can even be applied to learn continuous control fields.
		
The major bottleneck in using RL to control realistic experiments is set by sample efficiency. In the present implementation, I repeat each protocol $100$ times [not shown in Fig.~\ref{fig:reward_vs_episode}] to estimate its fidelity from the quantum measurement data. While this slows down the learning process, I stress that this is an intrinsic feature of all quantum measurements, unrelated to RL. In a large class of platforms, such as cold atoms, this can be alleviated by measuring multiple copies of the system simultaneously. Theoretically, the problem could also be mitigated by employing techniques from statistical inference, or a suitable pre-training procedure.
	
Even though the obtained fidelities are model-dependent and do not carry over to other control problems, the Q-Learning algorithm is universal in the sense that the RL agent, starting with no prior knowledge, learns only from its actions [protocols] and the stochastic reward. This is analogous to playing video or board games without seeing the game configuration, but only the (noisy) score. Hence, Q-Learning is agnostic on the fine details of the controlled system and can be applied to any model, even classical ones, as I demonstrate using the classical Kapitza pendulum, see \href{https://mgbukov.github.io/movies/RL_kapitza/movie-3.mp4}{Video 3} and App.~\ref{app:classical}.

\section{Outlook}

Every experiment comes with its own imperfections which obscure the physics of interest. Building a theory to describe them all in detail is often a formidable setup-dependent task, and requires considerable efforts. In the era of machine learning and automation, it is desirable to develop autonomous algorithms to delegate the tedious task of exploring the fine details of experiments to computers, and RL emerges as a natural candidate. It is currently an open question whether experimental imperfections can be turned into features, and exploited for the purpose of control.
	
In this respect, RL presents a set of promising algorithms, capable of simultaneously dealing with various sources of uncertainty and noise, even in highly complex far-from-equilibrium scenarios with no available analytical description. Out of a variety of RL algorithms~\cite{sutton1998reinforcement}, it is not clear which ones are best suited for controlling quantum systems away from equilibrium. In the current study, I chose Q-Learning, because it is \emph{off-policy}, i.e.~one can use data, generated when the policy of the agent was suboptimal, to improve the current policy. Further advantages are expected to be offered by Deep Learning~\cite{august_18,foesel_18,niu2018universal,huggins2018towards,carleo_17}, especially in the search of an efficient compressed representation of the state-action space, which is one way to incorporate continuous protocols. Deep Learning allows the agent to generalize and evaluate the value of previously unseen protocols, but also brings in difficulties associated with uncontrolled approximations and the absence of convergence guarantees for the algorithm. I verified that the tabular Q-Learning algorithm used in this paper is convergent.
	
This work represents a pioneering step in introducing RL to control quantum systems far away from equilibrium. Whereas it is difficult to a priori assess the suitability of RL for nonequilibrium many-body control, recent work successfully applied RL algorithms to control \emph{static} [i.e.~non-Floquet] chaotic many-body spin chains~\cite{bukov_17RL,august_18}. The higher complexity of many-body control may as well require to cast the RL problem as a partially-observable Markov decision process. While still at the beginning of this quest, the present proof-of-principle theoretical study already hints towards the applicability of RL to a large class of problems in quantum dynamics, and will hopefully spawn more research in this exciting new direction.

\emph{Acknowlegements.---}I wish to thank P.~Weinberg for help with speeding up the implementation of the Q-Learning algorithm, and M.~Stoudenmire for valuable discussions. This work was supported by the Emergent Phenomena in Quantum Systems initiative of the Gordon and Betty Moore Foundation, the ERC synergy grant UQUAM, and the U.S. Department of Energy, Office of Science, Office of Advanced Scientific Computing Research, Quantum Algorithm Teams Program. 
I used \href{https://github.com/weinbe58/QuSpin#quspin}{Quspin} for simulating the dynamics of the quantum and classical Kapitza oscillators~\cite{weinberg_17,quspin2}.
The author is pleased to acknowledge that the computational work reported on in this paper was performed on the Shared Computing Cluster which is administered by \href{https://www.bu.edu/tech/support/research/}{Boston University's Research Computing Services}.

\bibliographystyle{apsrev4-1}
\bibliography{./bibliography}

\clearpage

\appendix

\begin{widetext}

\section{\label{app:kapitza_theory}Simulating the Dynamics of the Kapitza Oscillator}

This section motivates the choice of Hamiltonian for the Kapitza oscillator [Eq.~\eqref{eq:kapitza_H}, main text], and discusses similarities and differences with the original Kapitza pendulum due to the periodic step drive.

The Kapitza oscillator with mass $m$ and natural frequency $\omega_0$ is governed by the Hamiltonian~\cite{bukov2015universal}
\begin{equation}
\label{eq:kapitza_lab}
H_\mathrm{lab}(t)=H_0 + A\Omega f(t)\cos\theta,\quad H_0=\frac{p_\theta^2}{2m} - m\omega_0^2\cos\theta,
\end{equation}
where $A$ and $\Omega=2\pi/T$ are the (dimensionless) amplitude and the frequency of the periodic drive, and $f(t+T)=f(t)$ is a $T$-periodic function with zero period-average. In the infinite-frequency limit, one may expect that, since the drive averages to zero, the system is effectively governed by the non-driven oscillator $H_0$. However, since the strength of the drive-system coupling scales linearly with the drive frequency $\Omega$, this na\"ive picture breaks down, and the effective infinite-frequency Floquet Hamiltonian $H_F(\Omega\to\infty)\neq H_0$.

In Ref.~\cite{bukov2015universal}, a generic way to circumvent this problem was suggested, by going to a rotating frame:
\begin{equation}
H_\mathrm{rot}(t) = V^\dagger(t)H_\mathrm{lab}(t)V(t) - iV^\dagger(t)\partial_t V(t), \quad V(t) = \exp\left(-i\Delta(t) \cos\theta \right), \quad \Delta(t)=A\Omega \int^t\mathrm{d}t' f(t').
\end{equation}
This transformation removes the linear with $\Omega$ scaling of the amplitude in  the time-integrated drive $\Delta(t)$, and facilitates the computation of the infinite-frequency Floquet Hamiltonian, as I now briefly revisit. Fundamentally, changing the reference frames can be seen as a resummation of an entire subseries of the inverse-frequency expansion~\cite{goldman2014periodically,bukov2015universal,eckardt2017atomic}. A straightforward calculation yields
\begin{equation}
H_\mathrm{rot}(t) = H_0 + \frac{1}{2m}\Delta(t)[\sin\theta, p_\theta ]_{+} - \frac{1}{2m}\Delta^2(t)\sin^2\theta,
\end{equation}
where $[\cdot,\cdot]_{+}$ denotes the anti-commutator. For instance, specializing to $f(t)=A\Omega\sin(\Omega t)$ gives $\Delta(t)=-A\cos(\Omega t)$ and $\Delta^2(t)=A^2/2(1+\cos2\Omega t)$. Since in this rotating frame the drive couplings (i.e.~the amplitudes) of $\Delta(t)$ and $\Delta^2(t)$ are independent of $\Omega$, the infinite-frequency Floquet Hamiltonian (up to a constant) can be computed by taking the time-average:
\begin{equation}
H_F(\Omega\to\infty) = H_0 - \frac{A^2}{8m}\cos 2\theta.
\end{equation}
Next to the free oscillator $H_0$, it contains an extra potential-energy term, which is responsible for stabilizing the inverted position of the oscillator at $\theta=\pi$ for high-enough frequencies. Note that this change of frames changes the micromotion (i.e.~intra-period) evolution, but not the Floquet Hamiltonian; hence, the stabilizing property of the dynamics is left intact.

To set up an efficient simulator for quantum dynamics which produces the data from which the RL agent learns, it is advantageous to consider periodic step-drives. These multi-harmonic analogues of the monochromatic drive allow to circumvent solving Schr\"odinger's equation using ODE integrators, and reduce simulation time. Typically, multi-harmonic drives do not change the structure of the Floquet Hamiltonian, i.e.~they preserve the stabilization effect: this can be seen with the help of the inverse-frequency expansion where the operator structure decouples from the time-ordered integrals and is, thus, independent of the specific choice of drive. However, there can be subtleties, which I now discuss.

A first guess would be to use a periodic-step time dependence for the lab-frame drive $f(t)$, with $\Delta(t)$ the corresponding continuous periodic zig-zag function. However, choosing a step-drive in the lab frame results in a more complicated time average $\int_0^T \Delta^2(t)\mathrm{d}t\sim\Omega^{-2}$, which eliminates the stabilizing $\cos2\theta$ term in the infinite-frequency Floquet Hamiltonian, and compromises the engineering property of the Floquet drive. 

This observation suggests to use step drives in the rotating frame. I propose the following periodic step-drive Hamiltonian:
\begin{equation}
H_\mathrm{rot}^\mathrm{step}(t) = H_0 - \frac{A}{2m}\mathrm{sign}\cos\Omega t\; [\sin\theta, p_\theta ]_{+} - \frac{A^2}{8m}\left(1-\mathrm{sign}\sin2\Omega t\right)\cos 2\theta.
\label{eq:Hrot}
\end{equation}
Certainly, Eq.~\eqref{eq:Hrot} has the correct infinite-frequency limit. However, the choice of the time-periodic step-functions in Eq.~\eqref{eq:Hrot} comes at a price, and a few remarks are in order:
(i) Notice that the time-dependence of the $\cos2\theta$-term is not equal to the square of the time-dependence in front of the $[\sin\theta, p_\theta ]_{+}$ term with this choice of drives. Hence, although the stroboscopic dynamics of the Hamiltonian~\eqref{eq:Hrot} at infinite-frequencies coincides with that of the original Kapitza oscillator~\eqref{eq:kapitza_lab}, this is not necessarily the case at finite frequencies. 
(ii) Even though the Hamiltonian in Eq.~\eqref{eq:Hrot} does not describe the original dynamics of the lab-frame oscillator, the Floquet Hamiltonian associated with Eq.~\eqref{eq:Hrot} also supports a stable equilibrium at high enough frequencies. Thus, the Floquet engineering properties of the Hamiltonian~\eqref{eq:Hrot} are the same as in the original Kapitza oscillator, although the corresponding finite-frequency Floquet Hamiltonians differ.
(iii) Sacrificing the equivalence of the micromotion dynamics by going from the lab to the rotating frame (and further by using step-drives), permits to also adjust the relative phases of the two time-dependent drives in Eq.~\eqref{eq:Hrot}.  The term $1/2\left(1-\mathrm{sign}\sin2\Omega t\right)$ allows to use a minimum of four (instead of eight) steps per driving cycle. Keeping the periodic drive commensurate with the bang-bang control further facilitates the simulation of the dynamics. In turn, this enables reaching longer evolution times at a small computational cost.
(iv) If one adds additional terms to the lab-frame Hamiltonian~\eqref{eq:kapitza_lab}, which depend on the position $\theta$ only, then they remain unaffected by the transformation to the rotating frame, and thus can simply be added to Eq.~\eqref{eq:Hrot}. Such a term is given, e.g., by the horizontal displacement operator $\sin\theta$, to which the control field couples [see main text]. Hence, Eq.~\eqref{eq:Hrot}  preserves the control properties of the original oscillator.

For these practical reasons, the simulator of quantum dynamics used to provide the data for the RL agent uses the Hamiltonian~\eqref{eq:Hrot}. I use the (angular) momentum basis with $21$ states in the Hilbert space, such that the low-energy initial and target wavefunctions remain marginally affected by increasing the number of states.

\section{\label{app:quant_meas}Simulating Quantum Measurements}

In this section, I motivate the specific choice of binary quantum measurement used to simulate a setup close to realistic experiments. Suppose one had access to all observables in the quantum Kapitza oscillator, and one could readily measure any combination of them. To determine if the state $\vert\psi(t_f)\rangle$ at the end of the protocol is the Floquet eigenstate localized at the inverted position, one would proceed as follows. Since one cannot measure states, but only observables, one has to measure the hermitian operator corresponding to the Floquet Hamiltonian $H_F$. A projective quantum measurement then returns probabilistically the $n$-th eigenvalue of $H_F$ with probability $\vert\langle\psi(t_f)\vert n_F\rangle\vert^2$, where $H_F\vert n_F\rangle=\varepsilon_F^n\vert n_F\rangle$. Let us reconcile this with the binary measurement defined in the main text: if, in a fixed outcome, $n$ coincides with the target state at the inverted position, this corresponds to the binary output $1$, and in all other cases -- to $0$.

Note that for many-body systems, the probability $\vert\langle\psi(t_f)\vert n_F\rangle\vert^2$ is likely to be exponentially small for almost all states, due to the exponentially large (with the system size) dimension of the Hilbert space. In such cases, it will be infeasible to successfully target a specific many-body state. This is related to the fact that the fidelity, being a probability, is a microscopic quantity, while in many-body systems the measurable quantities are observables (and their densities). For instance, in Ref.~\cite{bukov_17RL} it was demonstrated that in certain many-body control problems, one can successfully target microscopic states, such that the normalized logarithmic fidelity $-L^{-1}\log F_h(t_f)$ remains finite as $L\to\infty$. Another argument, based on typicality of many-body states~\cite{ETH_review}, shows that even though the eigenstates of generic observables are orthogonal by definition, within a small eigenvalue shell they share the same macroscopic properties, such as expectation values of observables, up to exponentially suppressed finite-size corrections. This raises the question whether it is possible to target macroscopic properties of many-body systems using RL and quantum control, to be addressed in future studies.

\section{\label{app:protocols}Control Protocols Learned by the RL Agent}

In this section I discuss the protocols learned by the RL agent. Even though I do not yet fully understand the physics behind the best RL protocols, certain features present themselves worthy of attention.

Figure~\ref{fig:protocols} (left column) shows the best protocol out of a family of $10^6$ local minima obtained using 1-flip SD, and the corresponding instantaneous fidelity evolution. The middle column shows the best encountered protocol by the RL agent during the train stage, according to the estimated fidelity [recall that the measurement is noisy and the agent only gets an estimate of the true value]. I checked that, for this seed realization, the agent in fact learned this protocol and was following it during the test stage. The right column shows the best true-fidelity protocol encountered during the train stage. Despite notable similarities at early times between the two protocols, there are small differences suggesting that the agent has hard time estimating the true value of a protocol close to optimality, due to the presence of noise in the reward. Additionally, notice that the instantaneous fidelities are not monotonic: they first rise and drop at intermediate times before they shoot up for the final values. In fact, the rise appears during a stage dominated by the $+4$ bang mode. This is reminiscent of the agent pushing the pendulum (on average) in one direction in order in the second stage to make use of the gained gravitational energy to overcome the potential barrier and eventually reach the inverted position from the other side in the time allotted. Indeed, this na\"ive classical picture is confirmed by the protocol visualizations [see \href{https://mgbukov.github.io/movies/RL_kapitza/movie-1.mp4}{Video 1}]. The quantum nature of the dynamics is most likely hidden in the non-trivial character of the bang-sequence.  

\begin{figure} 
	
	\begin{tabular}{ll}
		\includegraphics[width=0.333\columnwidth]{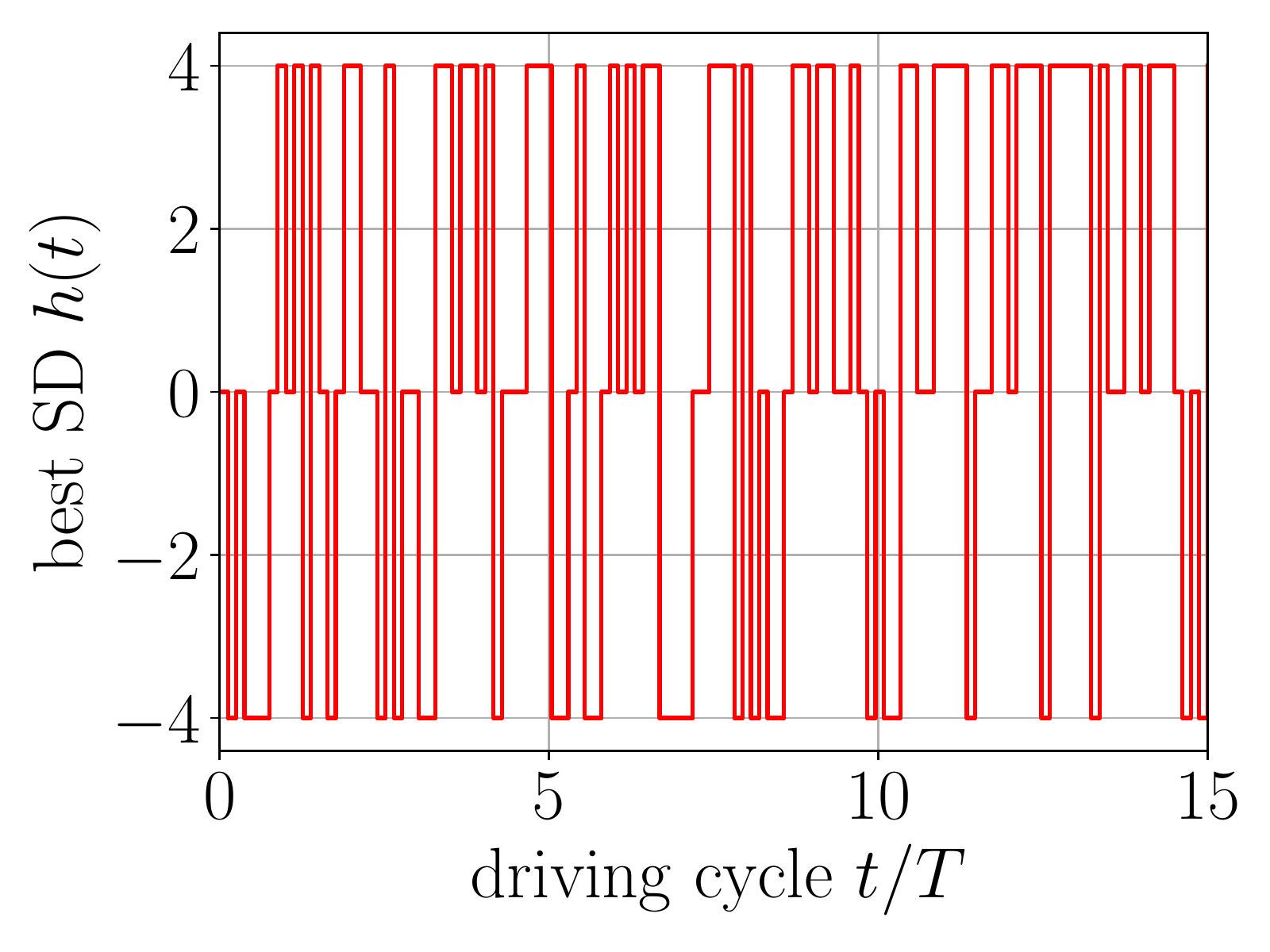}
		\includegraphics[width=0.333\columnwidth]{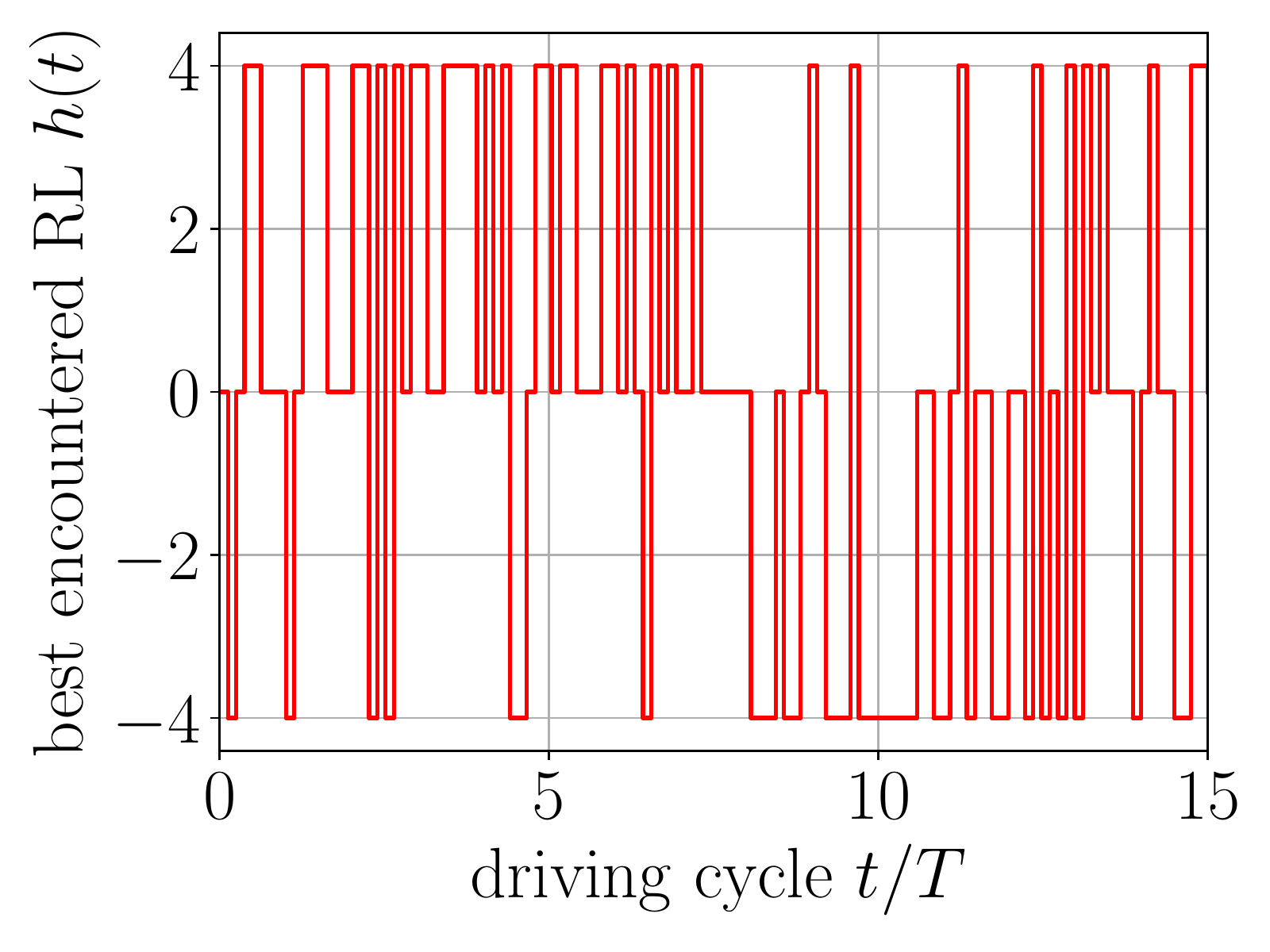}
		\includegraphics[width=0.333\columnwidth]{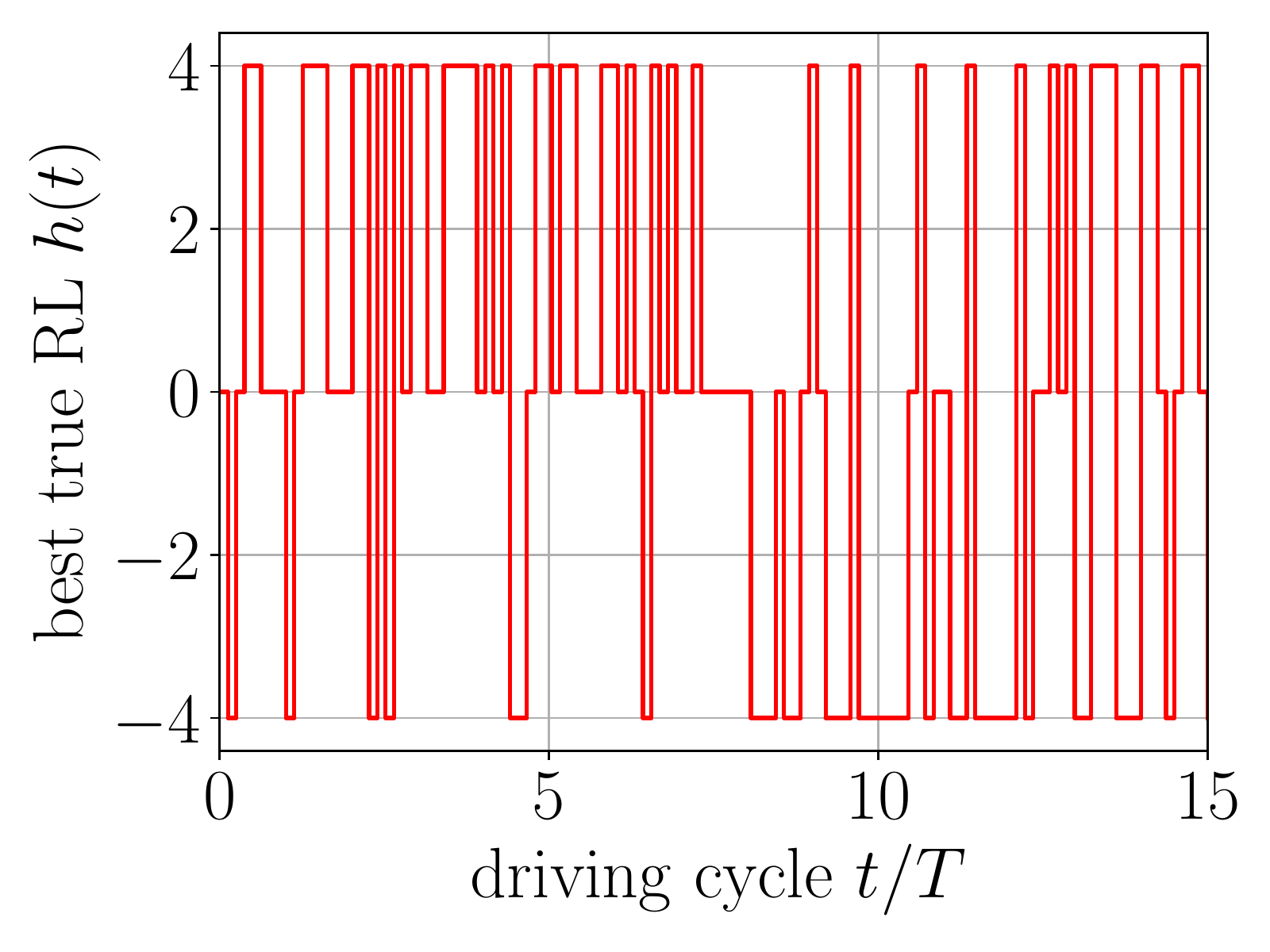}		
	\end{tabular}

	\begin{tabular}{ll}
		\includegraphics[width=0.333\columnwidth]{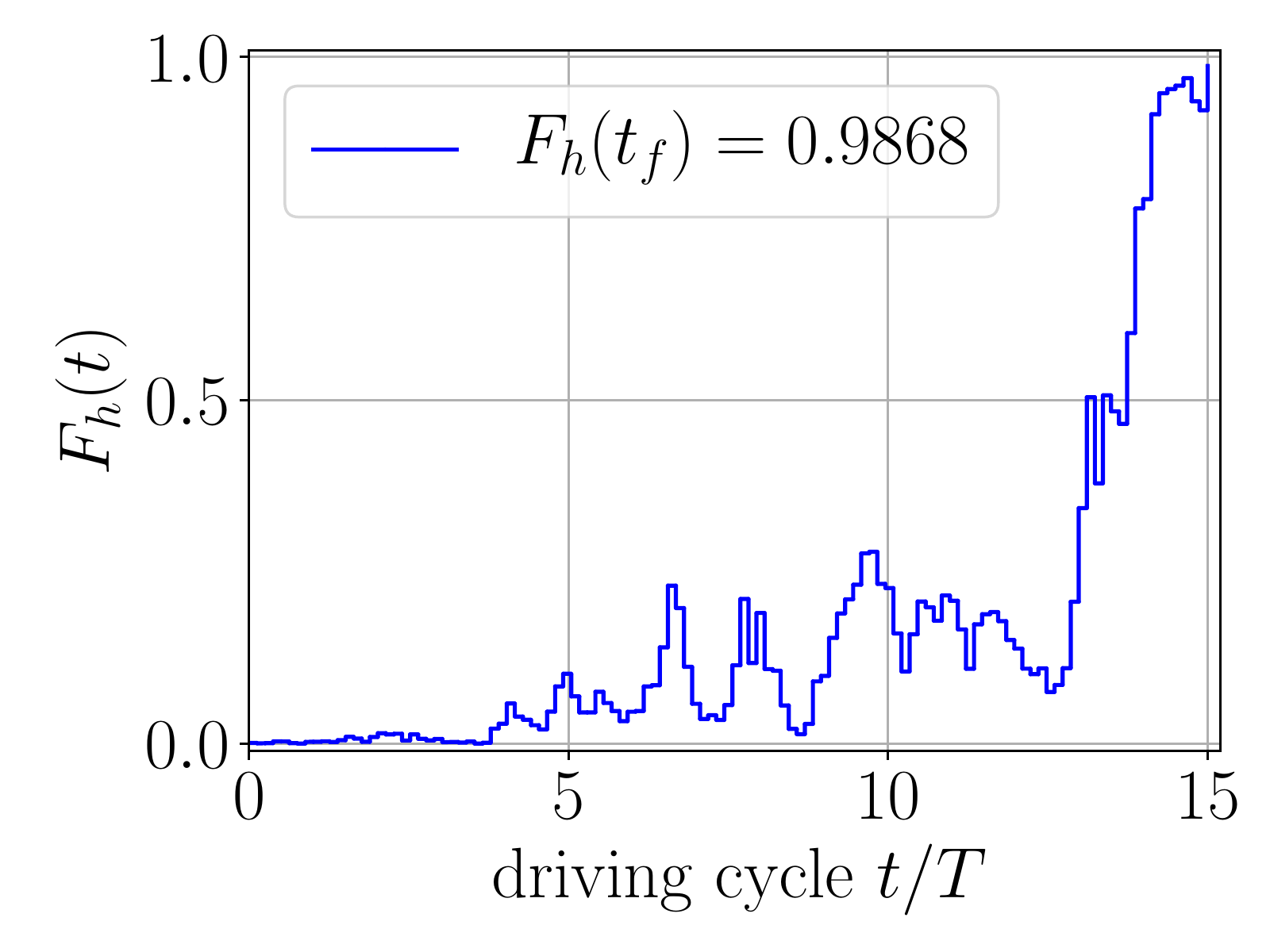}
		\includegraphics[width=0.333\columnwidth]{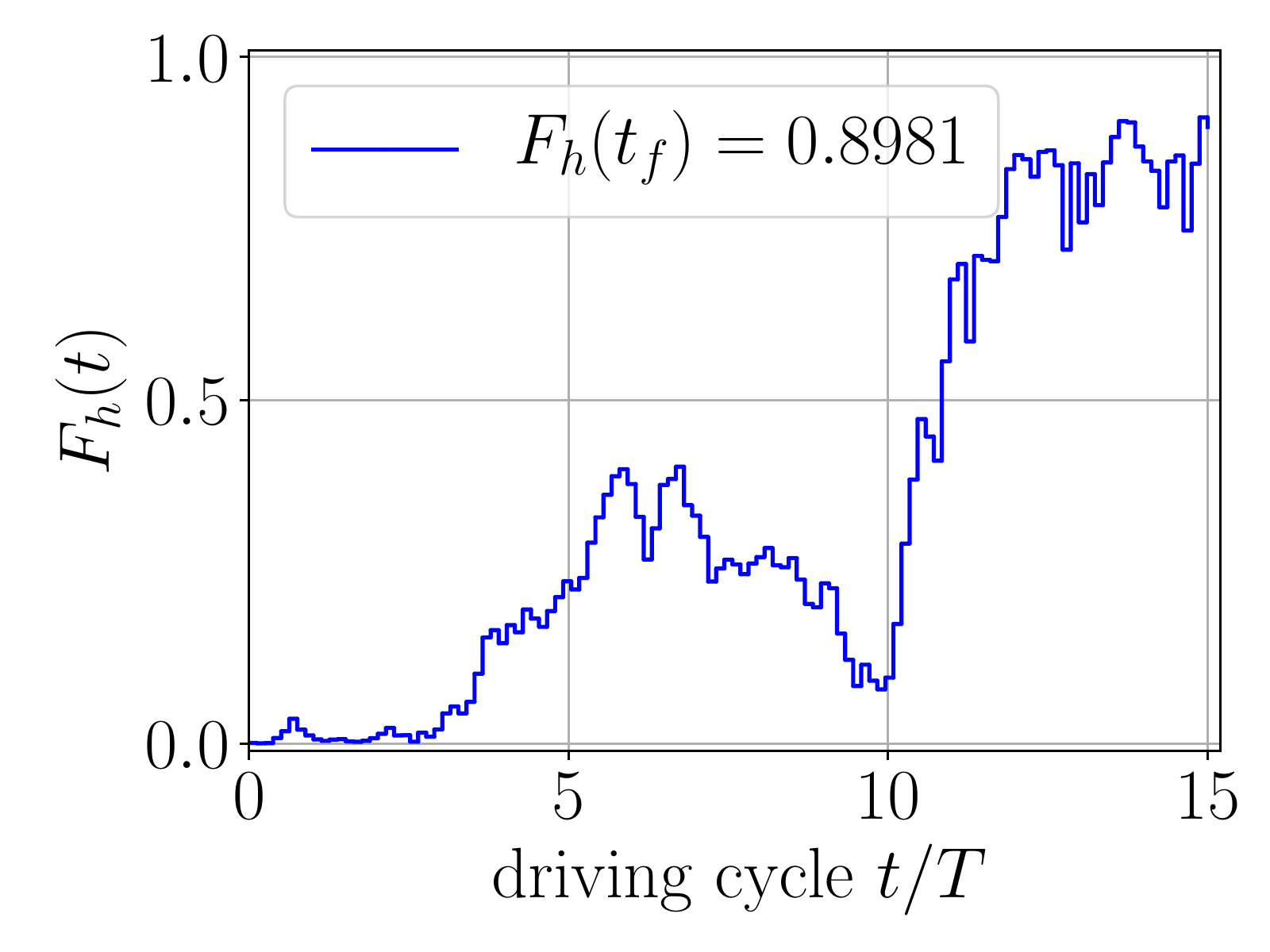}
		\includegraphics[width=0.333\columnwidth]{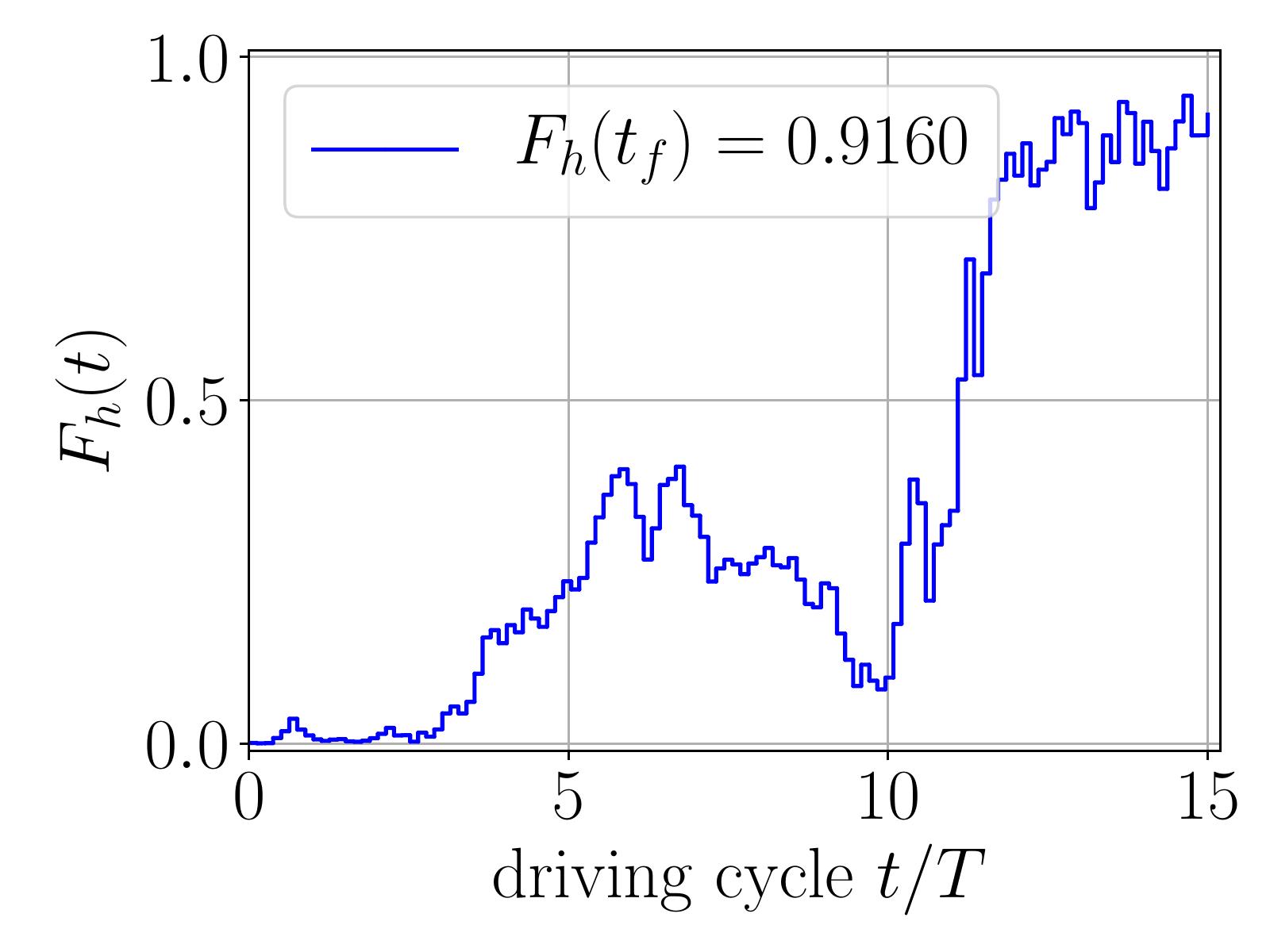}		
	\end{tabular}
	
	\caption{\label{fig:protocols}Protocols (upper row) and the corresponding fidelities (lower row) against time. The oscillator parameters are $\Omega/\omega_0=10$, $A=2$ and $m\omega_0=1$. The protocols contain $8$ steps per period for a total of $N_T=15$ periods.} 
	
\end{figure}

While it is hard to make precise sense of these protocol patterns, one can gain insights into the complexity of Kapitza bang-bang control as an optimization problem. Figure~\ref{fig:fid_hist} (left panel) shows the histogram of $10^6$ randomly chosen protocols. As observed using the train curves in RL, the mean fidelity of a random protocol is about $10\%$, which is consistent. This distribution represents the density of states (DOS) in protocol space~\cite{day2018glassy}. It decays at least exponentially with fidelity (middle panel). This distribution comes in strong contrast to that of 1-flip local SD minima in the infidelity landscape (right panel), which is heavily sifted towards the high-fidelities of interest. When pushed to $10^6$ training episodes, the RL agent learns on average a fidelity which is consistent with the mean of this distribution, cf.~Fig.~\ref{fig:fid_traces_vs_Nep} [top right panel].

\begin{figure} 
	
	\begin{tabular}{ll}
		\includegraphics[width=0.333\columnwidth]{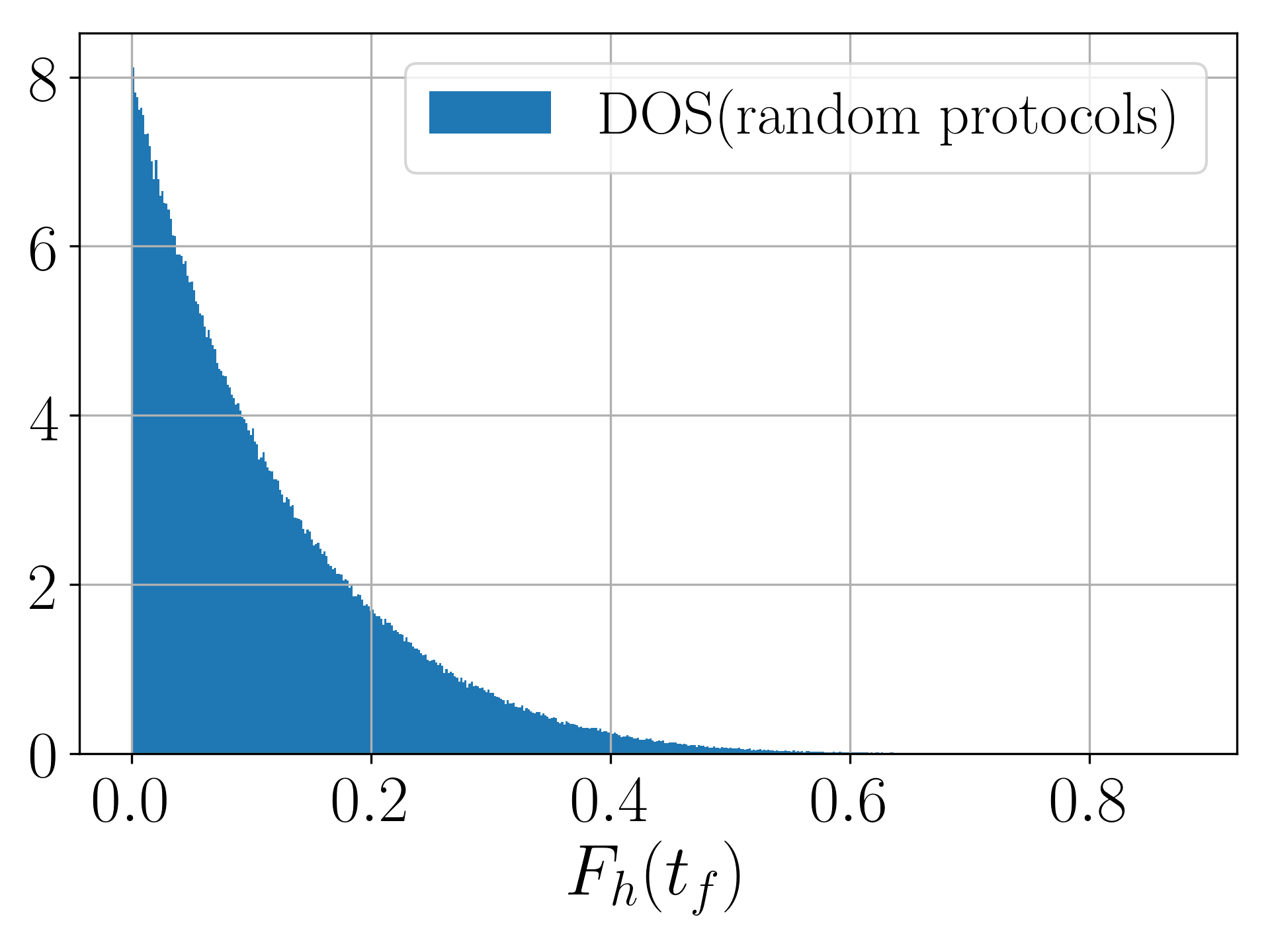}
		\includegraphics[width=0.333\columnwidth]{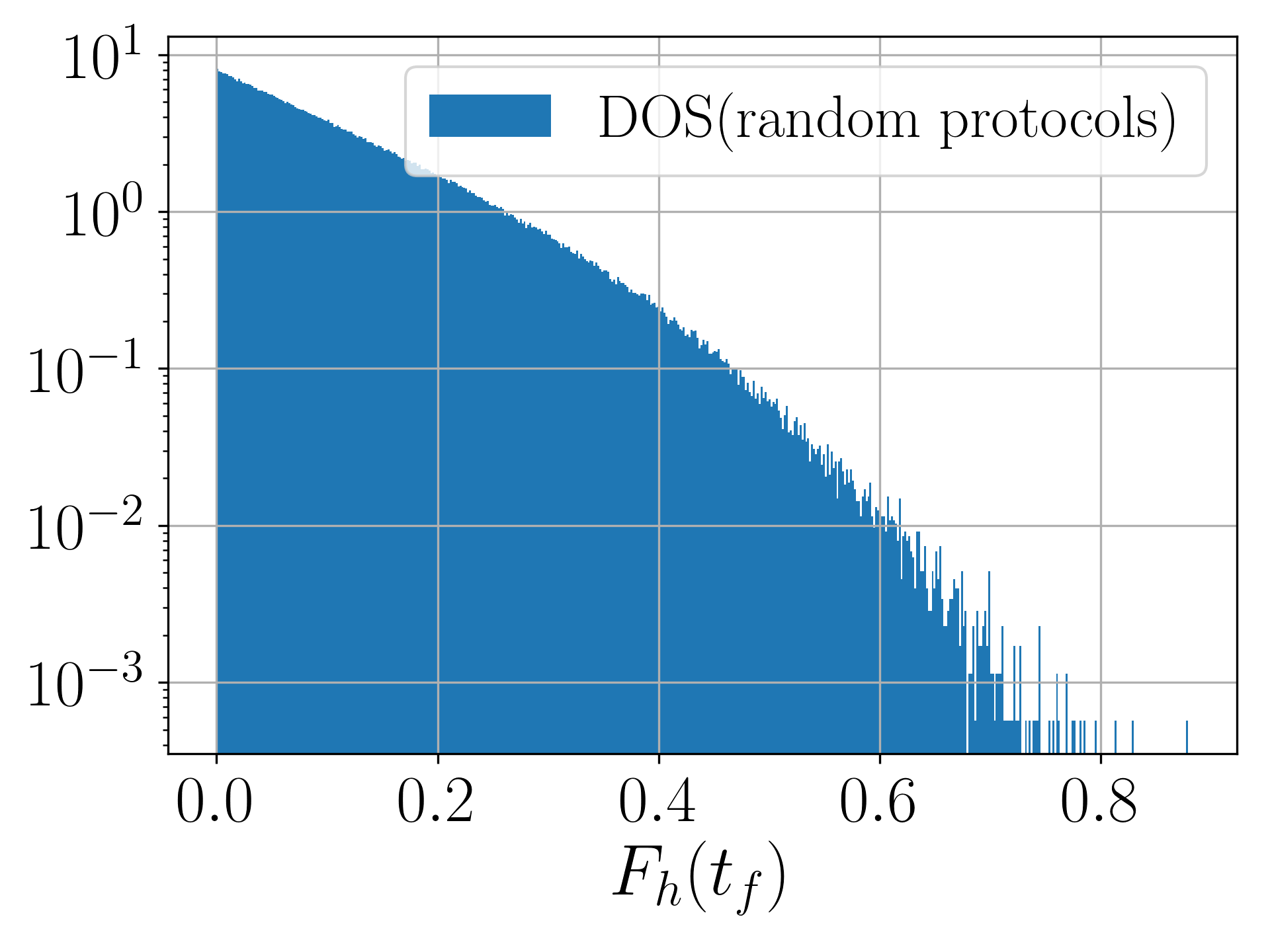}
		\includegraphics[width=0.333\columnwidth]{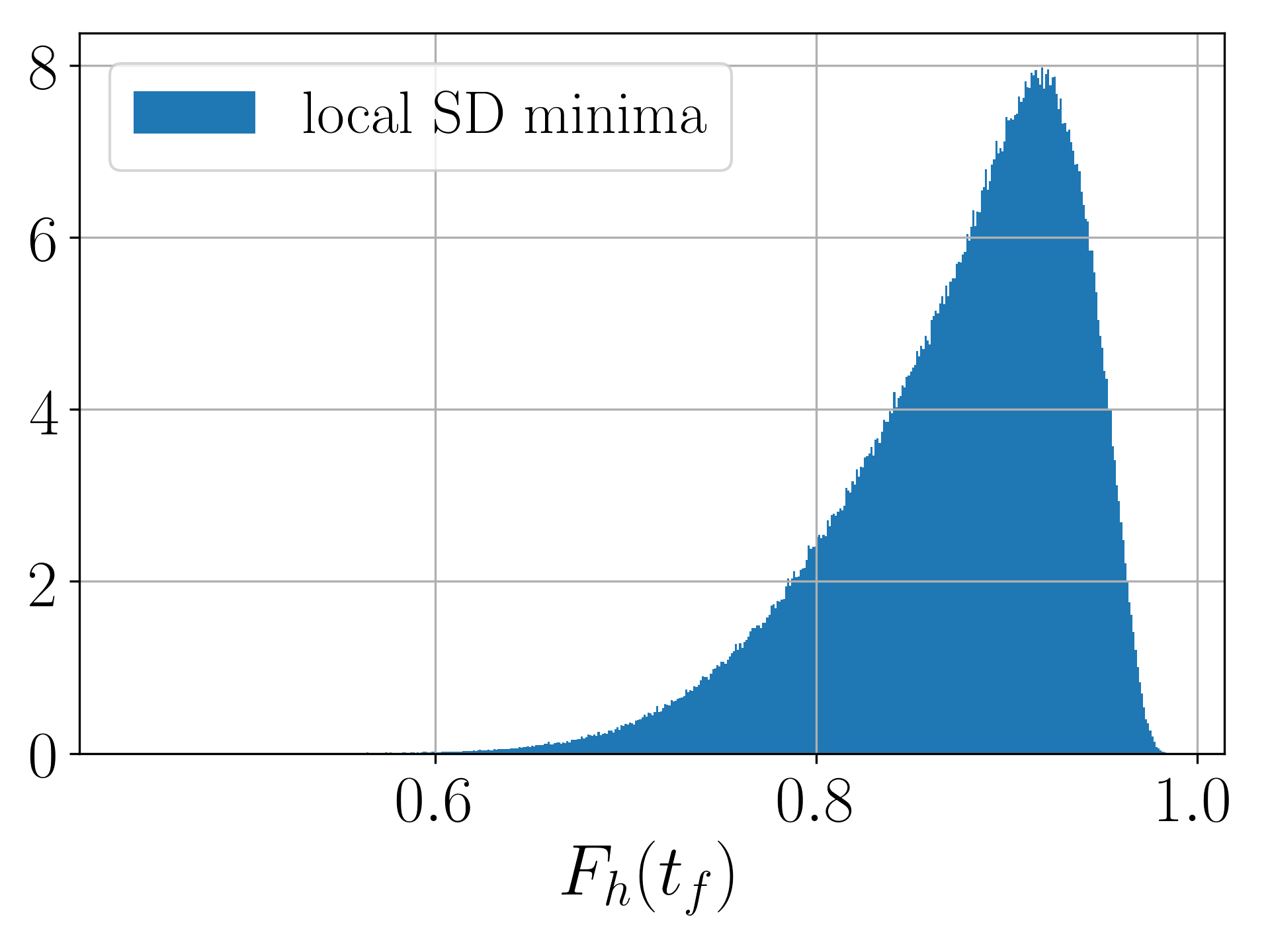}		
	\end{tabular}

	\caption{\label{fig:fid_hist}Fidelity distributions for random protocols on linear (left), and a semi-log (middle) scale, and the sample of local SD-minima (right). The oscillator parameters are $\Omega/\omega_0=10$, $A=2$ and $m\omega_0=1$. The protocols contain $8$ steps per period for a total of $N_T=15$ periods.}

\end{figure}

\section{\label{app:gaussian_target}Learning a Quasi-Gaussian Target State}

As I explained in Sec.~\ref{app:quant_meas}, targeting an exact Floquet eigenstate requires the ability to measure the Floquet Hamiltonian. Unfortunately, often this is beyond the capabilities of present-day experiments. 

\begin{figure}[t!]
	\includegraphics[width=0.5\columnwidth]{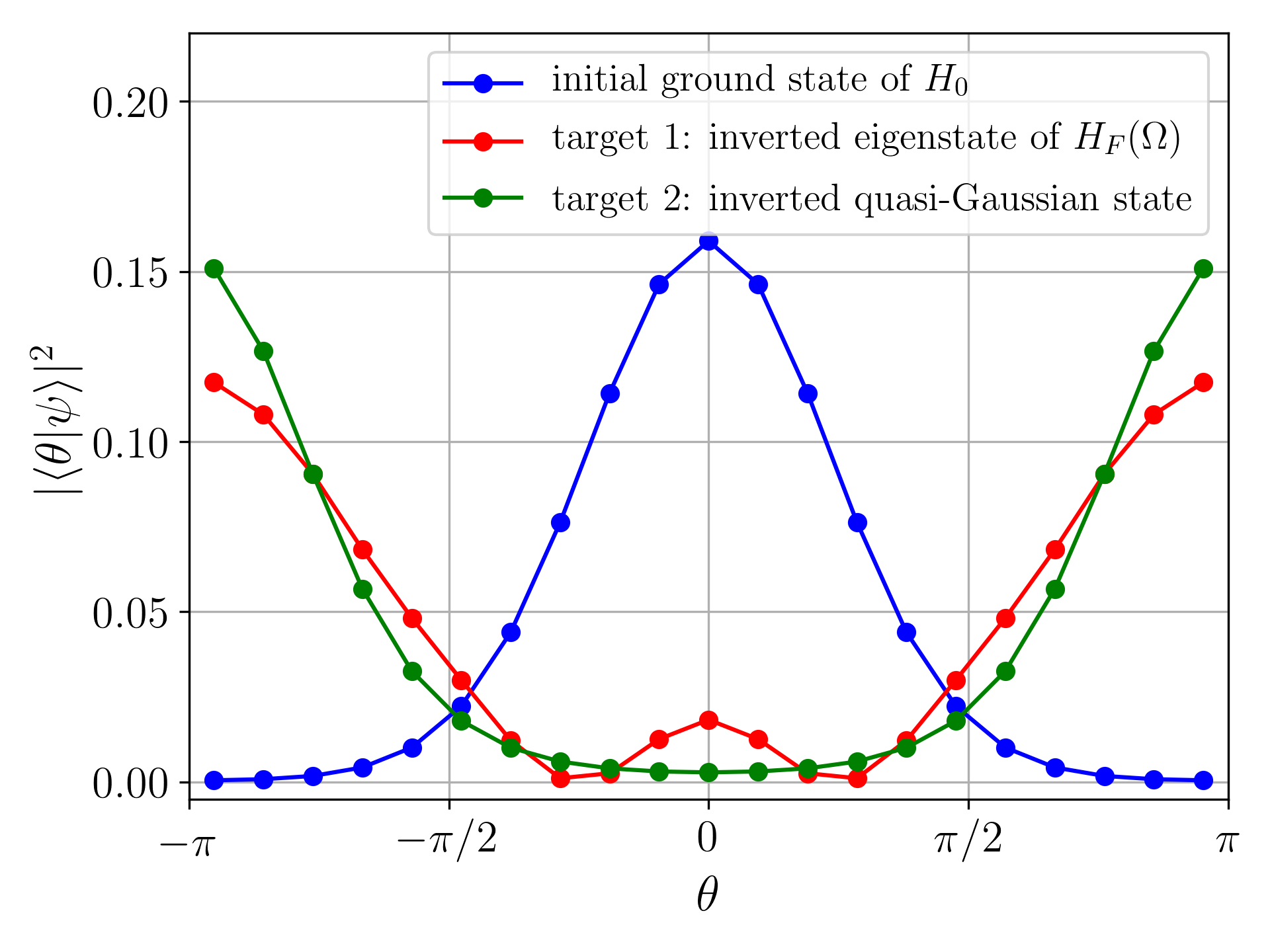}
	\caption{\label{fig:quantum_states} Initial state and target states at the inverted position for $\Omega/\omega_0=10$, $A=2$ and $m\omega_0=1$.}
\end{figure}

At the same time, however, the main reason behind the interest in that particular Floquet eigenstate, is precisely its feature to be localized at the inverted position. One might then wonder how the RL agent would perform, if required to target a simpler state with the same property. To test this, I use as an alternative target state the quasi-Gaussian state $\langle\theta|\psi_\ast\rangle\!\propto\!\exp(-a^{-1/4}\cos\theta)$ [Fig.~\ref{fig:quantum_states}], with oscillator length $a=\left(\omega'\right)^{-1/4}$ in the harmonic approximation set by the infinite-frequency effective potential at the inverted position: $\omega'\!=\!\sqrt{A^2/(2m^2) \!-\! \omega_0^2}$. Such a state can be emulated easily as the ground state of some static Hamiltonian, and is thus much more easily accessible compared to the exact Floquet eigenstate. Figure~\ref{fig:reward_vs_episode_gauss} shows that this does not introduce any additional difficulties for the RL algorithm. As a matter of fact, the obtained fidelities are slightly higher, compared to targeting the exact Floquet eigenstate, cf.~Fig.~\ref{fig:reward_vs_episode} in the main text.

\begin{figure}
	\includegraphics[width=0.6\columnwidth]{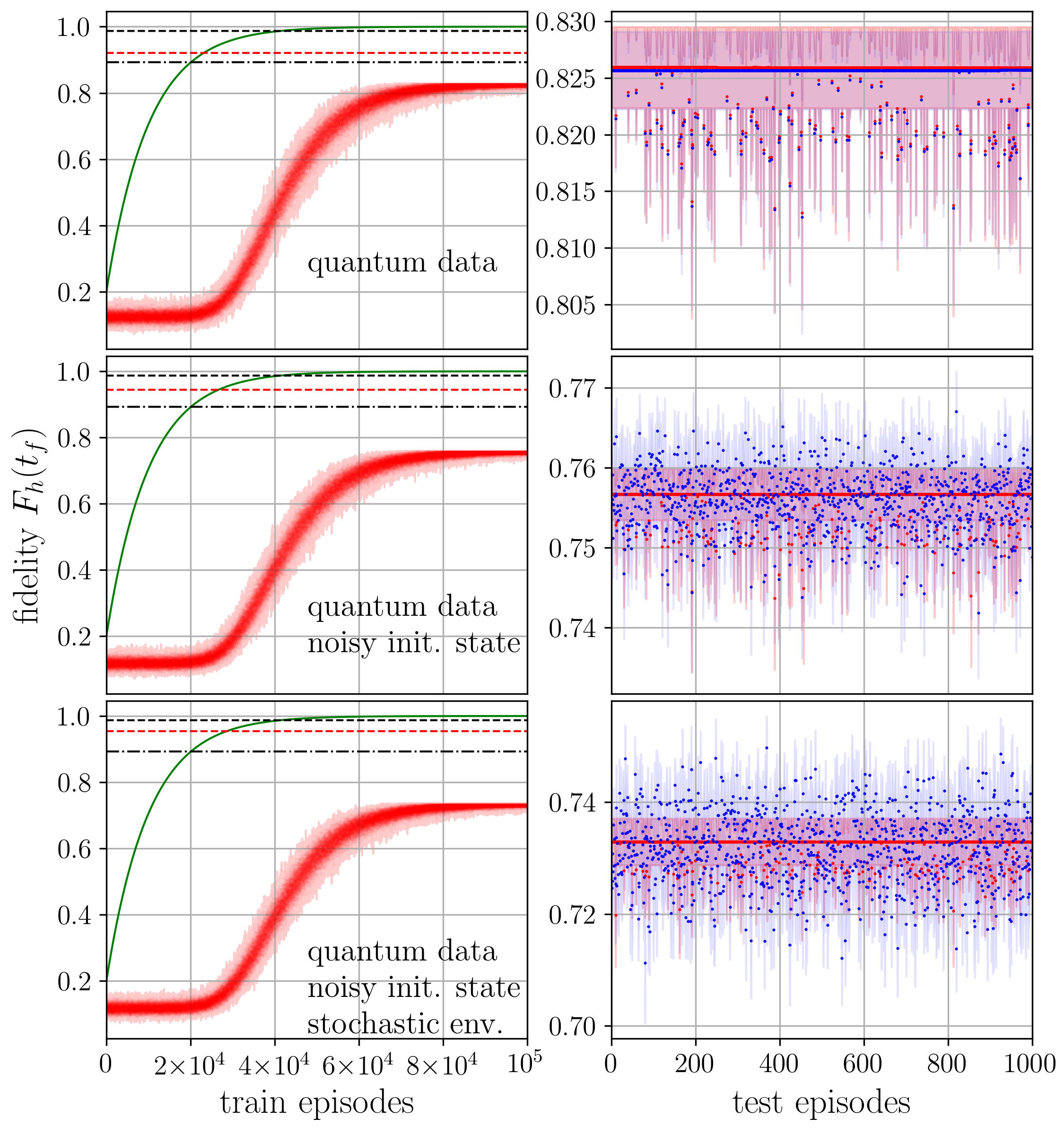}
	\caption{\label{fig:reward_vs_episode_gauss}Train and test stages for the quasi-Gaussian target state $\langle\theta|\psi_\ast\rangle\!\propto\!\exp(-a^{-1/4}\cos^2\theta)$, localized at the inverted position. The details are the same as in Fig.~\ref{fig:reward_vs_episode}, main text.}
\end{figure}

\section{\label{app:learning}Q-Learning in Noisy and Stochastic Environments}

\subsection{\label{subapp:noise_exp}Numerical Experiment: does the Q-Learning Agent Learn Specifics of the Stochastic Environment?}

Training the Q-Learning agent with noise in the initial state and in a stochastic environment, raises the question whether it is capable of learning the details of such uncertainty sources and exploit them to its advantage in the learning process. To this end, I consider the following numerical experiment: the agent is trained on a noise-free deterministic environment [with the only uncertainty in the reward, as a result of the quantum measurement], but subsequently tested on a noisy stochastic environment [i.e.~with additional occasional random failures in the bangs of the protocols]. The performance during the test stage should then be compared to the case where the agent was also trained in a noisy stochastic environment [$\eta=0.31$, $\zeta=1/120$, see main text]. In which scenario does the agent perform better?

To answer this, I distinguish between two quantities during the test stage: the estimated fidelity expected by the agent (red) and the true fidelity of the protocol (blue). Figure~\ref{fig:exp_reward_vs_episode} clearly shows that, after learning in a noise-free deterministic environment, the agent erroneously learns to expect a higher fidelity, compared to the true fidelity associated with the learned protocol. 

There are two important conclusions from this numerical experiment. (i) Notice that the true fidelity in the train stage in Fig.~\ref{fig:exp_reward_vs_episode} (blue dots) is about the same as the expected fidelity had the agent been trained in the presence of uncertainty [see Fig.~\ref{fig:reward_vs_episode} (red dots), main text]. This suggests that the agent does not learn to exploit any additional features of the environment using this state-action space parametrization. This means that the RL algorithm is intrinsically robust to noise. The most likely reason for this lies in the stochastic $\varepsilon$-greedy exploration schedule used in Q-Learning, cf.~Sec.~\ref{app:algo}. (ii) Recall that the RL state space definition depends on the initial state $\vert\psi_i\rangle$. As the noise induces a change in the initial state, the algorithm learns the average fidelity over an ensemble of noisy initial states. This reveals the weakness of the current state-action choice to generalize to arbitrary initial conditions, which comes as a trade-off to the capability to learn solely from the actions taken. This behavior can be explained by the observation that the agent is not presented with any information about the uncertainty in the environment during learning: e.g., if the agent was told retroactively once a bang in a protocol had randomly failed, it might be possible to learn to `correct' or `counteract' this change. This behavior will be explored in future studies.

\begin{figure}
	\includegraphics[width=0.7\columnwidth]{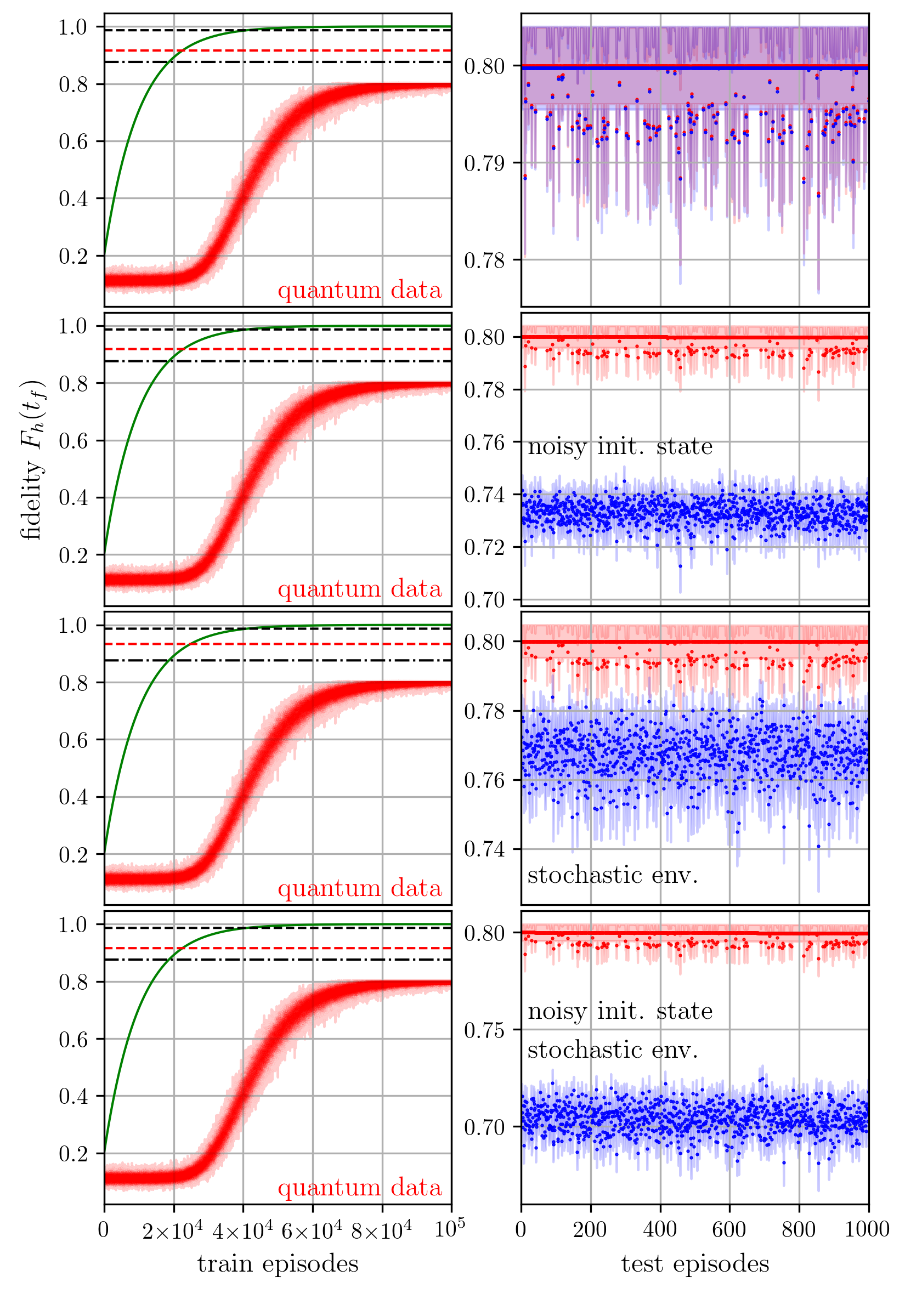}
	\caption{\label{fig:exp_reward_vs_episode}Numerical experiment training the agent using binary quantum data in a deterministic environment starting from a fixed initial state, while testing on noisy initial states and/or stochastic environment. The red data shows the agent's estimated fidelity, while the true fidelity is shown in blue. The mismatch between the agent's estimate (red) and the true fidelity (blue) arises due to the absence of noise and stochasticity during the Train stage. The left side (train curves) shows the same data for better comparison. The parameters are the same as in Fig.~\ref{fig:reward_vs_episode} of the main text.}
\end{figure}

\subsection{\label{subapp:single_vs_average}Single Run vs.~Average Performance}

Even in deterministic setups, the Q-Learning algorithm, cf.~Sec.~\ref{app:algo}, contains intrinsic noise due to the $\varepsilon$-greedy exploration schedule used during the train stage. Therefore, the algorithm is run for $100$ independent realizations of the pseudo-random number generator, and the graphs in this paper show averages. The deviation from the mean is computed using a bootstrapping approach (shaded area). Figure~\ref{fig:fid_traces_single-shot} shows the worst (left) and the best (middle) runs, and compares them to the average fidelity performance (middle).

\begin{figure}[h!]
	\includegraphics[width=1.0\columnwidth]{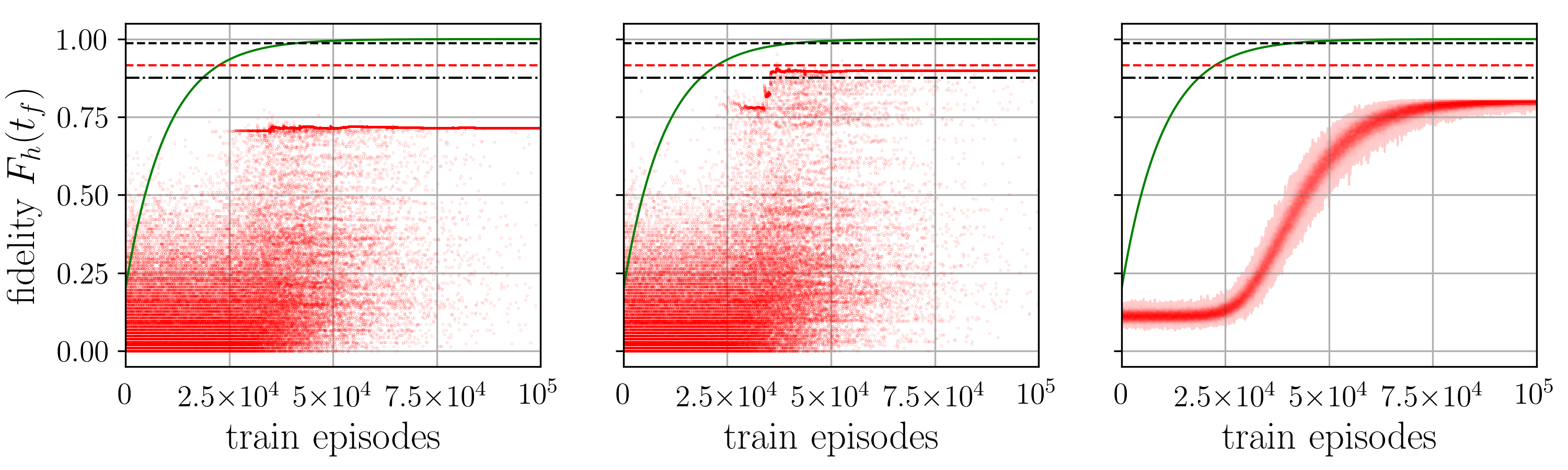}
	\caption{\label{fig:fid_traces_single-shot}Training behavior as a function of the number of Train episodes: worst run (let), best run (middle) and average performance (right) over $100$ seed realizations of the pseudo-random number generator. The oscillator parameters are $\Omega/\omega_0=10$, $A=2$ and $m\omega_0=1$. The protocols contain $8$ steps per period for a total of $N_T=15$ periods. The target is the Floquet eigenstate. The initial state is noise-free and the environment is deterministic.}	
\end{figure}

\subsection{\label{subapp:episode_scaling}Dependence on the Number of Training Episodes}

It is curious to study how the agent's learning capabilities change as the number of training episodes increases. As noted in the main text, the huge protocol space contains $3^{120}\sim 10^{57}$ protocol configurations. Additionally, the fidelity histograms, cf.~Fig.~\ref{fig:fid_hist} (left panel), show that most states have very poor fidelities. In the main text I showed data for up to $10^5$ training episodes. Figure~\ref{fig:fid_traces_vs_Nep} shows that one can achieve a reasonable improvement by increasing the training episodes by an order of magnitude. I do not consider it appropriate to push the Q-Learning algorithm to its limits, since the maximum number of training episodes in realistic experimental setups is set by the sample efficiency. Instead, I believe that the algorithm can be made more useful if it is appropriately improved in sample efficiency instead. 

\begin{figure}
	
	\begin{tabular}{ll}
		\includegraphics[width=0.333\columnwidth]{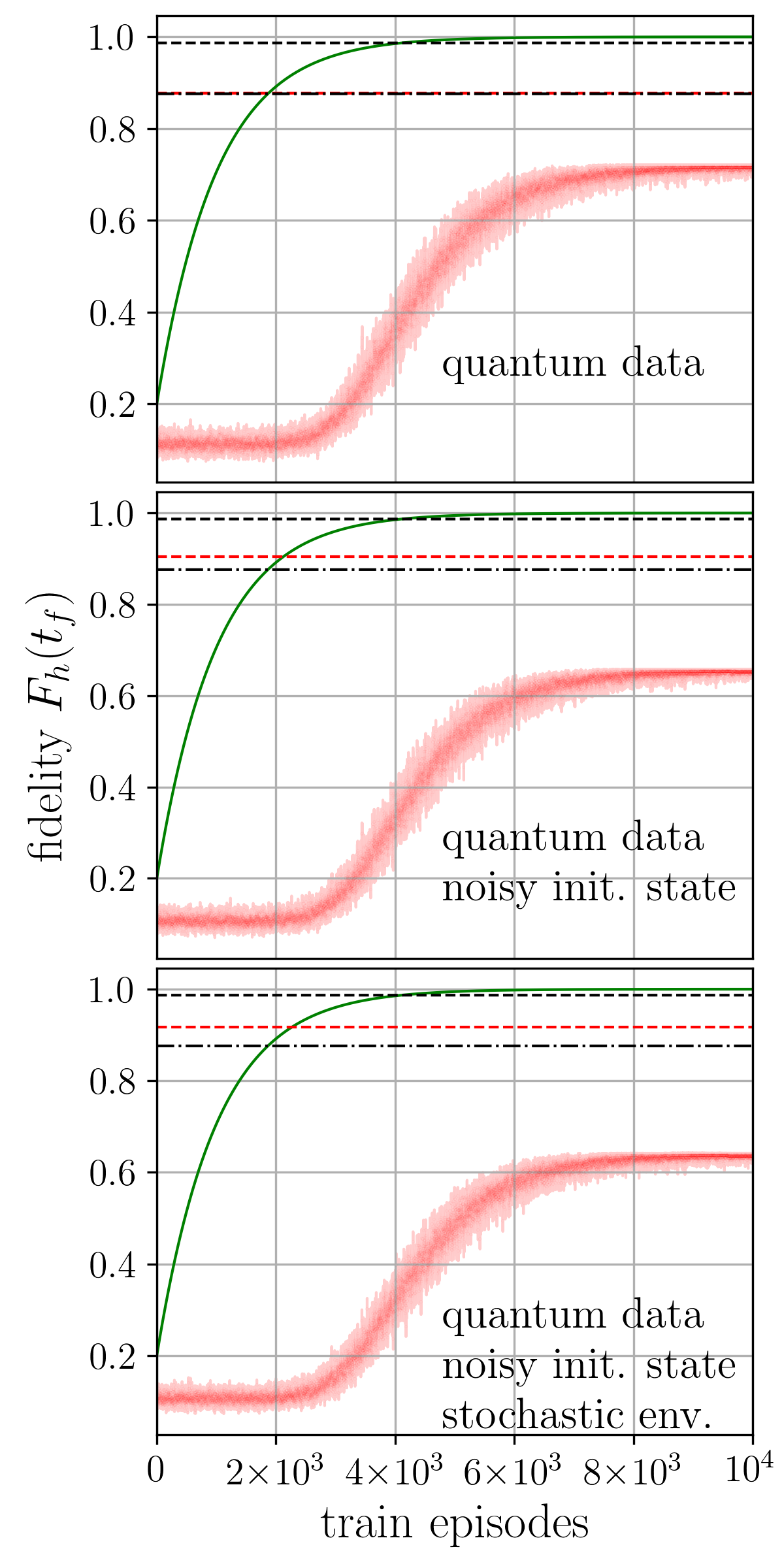}
		\includegraphics[width=0.333\columnwidth]{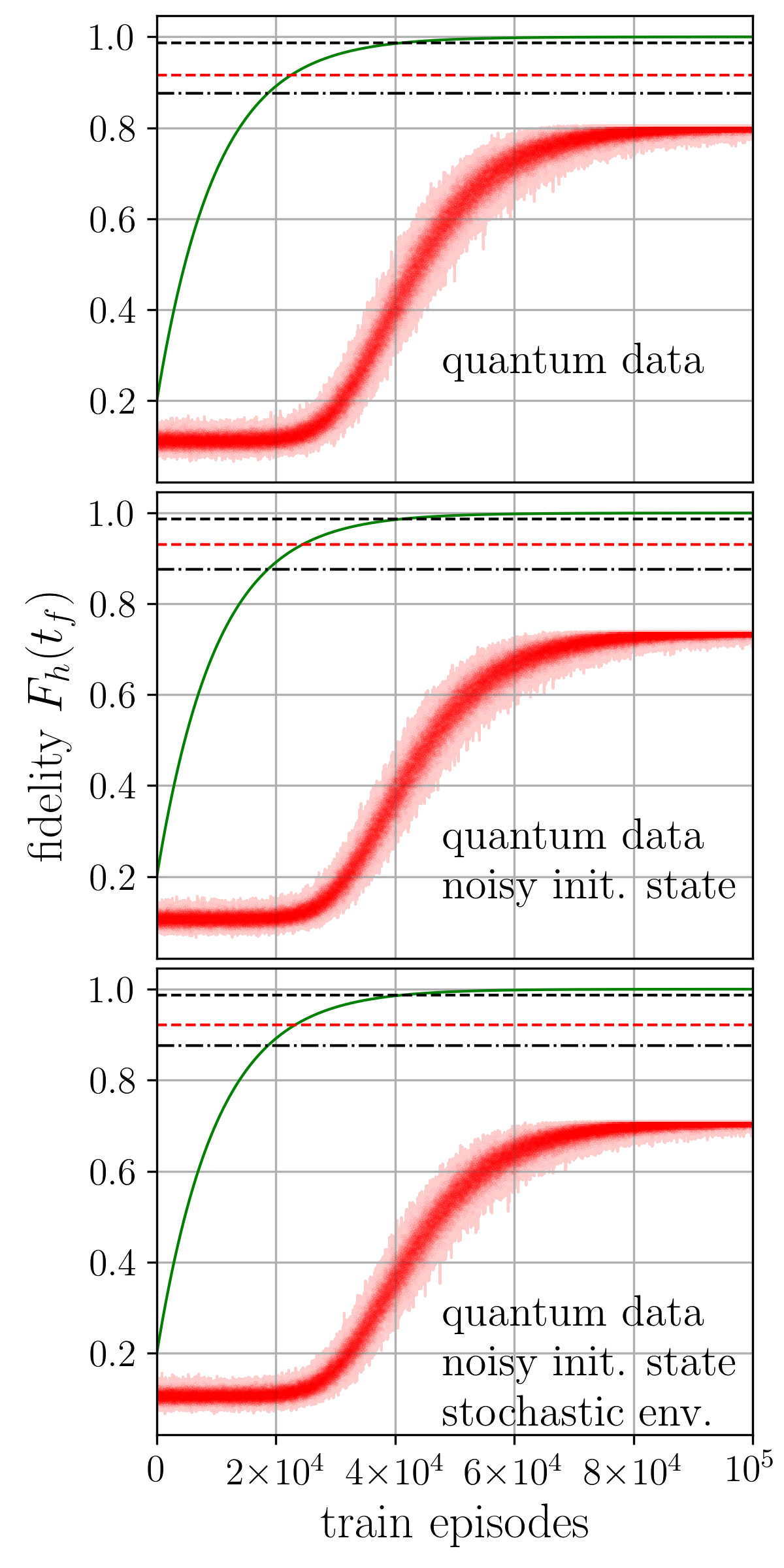}
		\includegraphics[width=0.333\columnwidth]{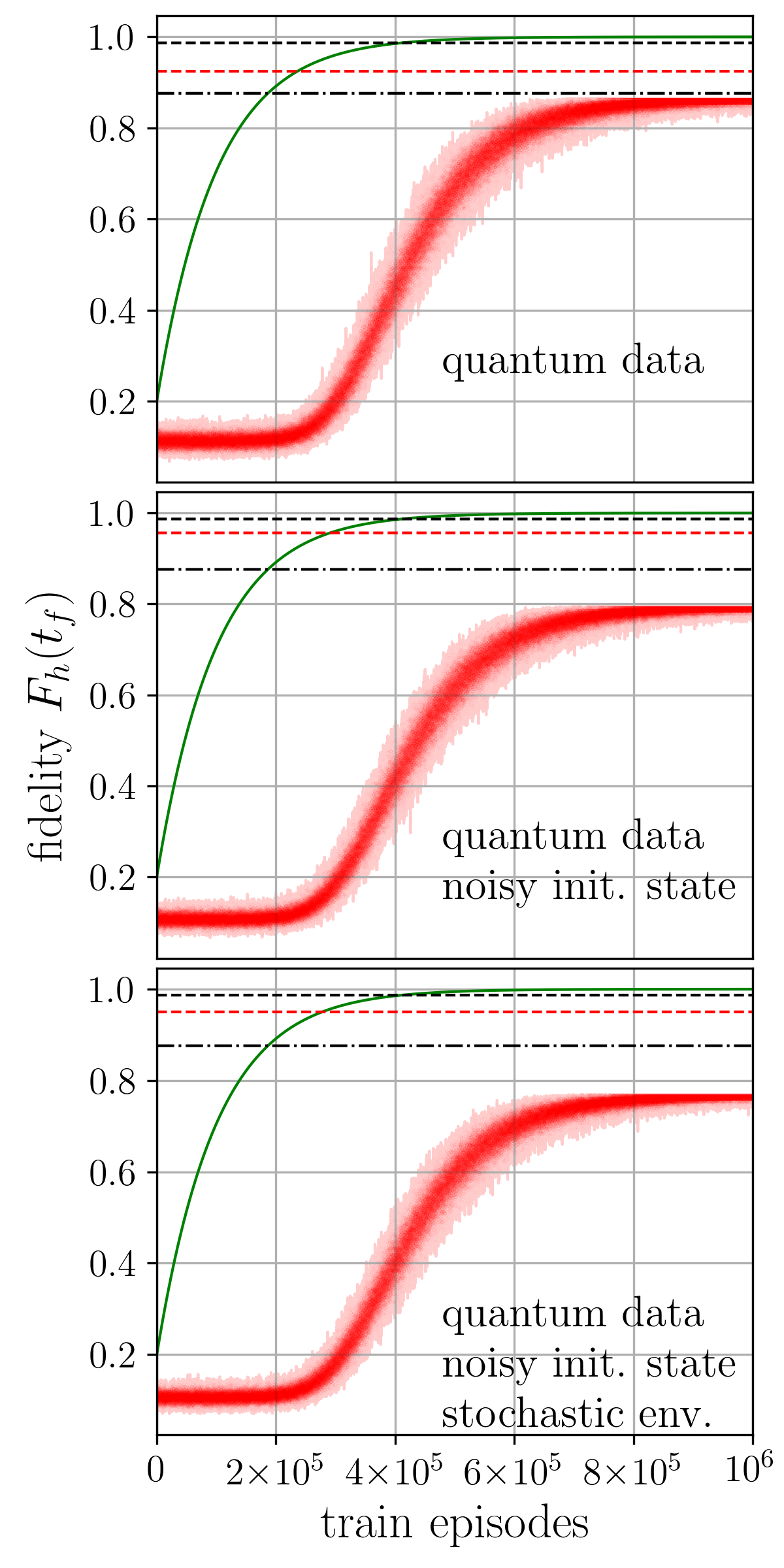}		
	\end{tabular}
	
	\caption{\label{fig:fid_traces_vs_Nep}Training behavior as a function of the number of Train episodes: $10^4$(left), $10^5$ (middle), and $10^6$(right). The oscillator parameters are $\Omega/\omega_0=10$, $A=2$ and $m\omega_0=1$. The protocols contain $8$ steps per period for a total of $N_T=15$ periods. The target is the Floquet eigenstate.}
	
\end{figure}

\section{\label{app:classical}Reinforcement Learning to Invert the Classical Kapitza Pendulum}

Last but not least, I demonstrate the versatility and universality of the Q-Learning algorithm by applying it to the classical Kapitza pendulum. For the sake of a better comparison with the quantum Kapitza oscillator, I choose the same time-dependent Hamiltonian in the rotating frame:
\begin{equation}
H(t) = \frac{p_\theta^2}{2m} - m\omega_0^2\cos\theta - \frac{A}{m}\left(\mathrm{sign}\cos\Omega t\right) p_\theta\sin\theta - \frac{A^2}{8m}\left(1-\mathrm{sign}\sin2\Omega t\right)\cos 2\theta + h(t)\sin\theta,
\label{eq:Hrot_cl}
\end{equation}
where $p_\theta$ and $\theta$ are classical conjugate variables, and $h(t)$ is the bang-bang control field. The dynamics of the pendulum is governed by Hamilton's equations of motion:
\begin{eqnarray}
\label{eq:Hamiltons_EOM}
\dot \theta  &=& \frac{1}{m}\left( p_\theta - A\left(\mathrm{sign}\cos\Omega t\right) \sin\theta \right), \nonumber\\
\dot p_\theta &=& -m\omega_0^2 \sin\theta + \frac{A}{m}\left(\mathrm{sign}\cos\Omega t\right) p_\theta\cos\theta
- \frac{A^2}{4m}\left(1-\mathrm{sign}\sin2\Omega t\right)\sin 2\theta - h(t)\cos\theta.
\end{eqnarray}
The initial state is chosen as $\theta(t=0)=0.01$ and $p_\theta(t=0)=0$. The target state is the inverted position at $\theta=\pi$. The finite value of the initial angle breaks the symmetry of the optimal protocol [i.e.~reaching the target clockwise and counter-clockwise]. 

To define the reward for the agent, note that simply reaching the target is not enough to assure a stable orbit at the inverted position after the control sequence is over, if the momentum at the end of the protocol is large enough to cause spin-over. Hence, a good cost function should penalize large final momenta. I thus (empirically) choose the following reward:
\begin{equation}
r(\theta,p_\theta) = \frac{1}{\pi^2}\left[ (\theta(t_f)+\pi)\mathrm{mod}(2\pi) - \pi \right]^2 - 4\vert p_\theta(t_f)\vert^2,\qquad r\in[-\infty ,1].
\label{eq:classical_reward}
\end{equation} 

In classical systems, measurements are deterministic. However, they might still be noisy. To take this into account in RL, I add Gaussian noise to the values of the position and momentum with zero mean and variance $\sigma=0.05$. This leads to uncertain rewards. Similar to the quantum case, I also consider cases in which (i) the initial state is noisy, by adding Gaussian noise in the initial condition with zero mean and variance $\eta=0.1$, and (ii) there are occasional failure events in the control bangs. This works in the same way as for the quantum oscillator. 

The universality of the state-action representation makes the Q-Learning algorithm agnostic on the physical system it is applied to. Thus, I apply the same algorithm to the classical Kapitza pendulum, see Fig.~\ref{fig:fid_traces_classical}. The best RL protocol for the case of noisy reward but noiseless initial state in a deterministic environment is shown in \href{https://mgbukov.github.io/movies/RL_kapitza/movie-3.mp4}{Video 3}. 

\begin{figure}[h!]
	\includegraphics[width=0.6\columnwidth]{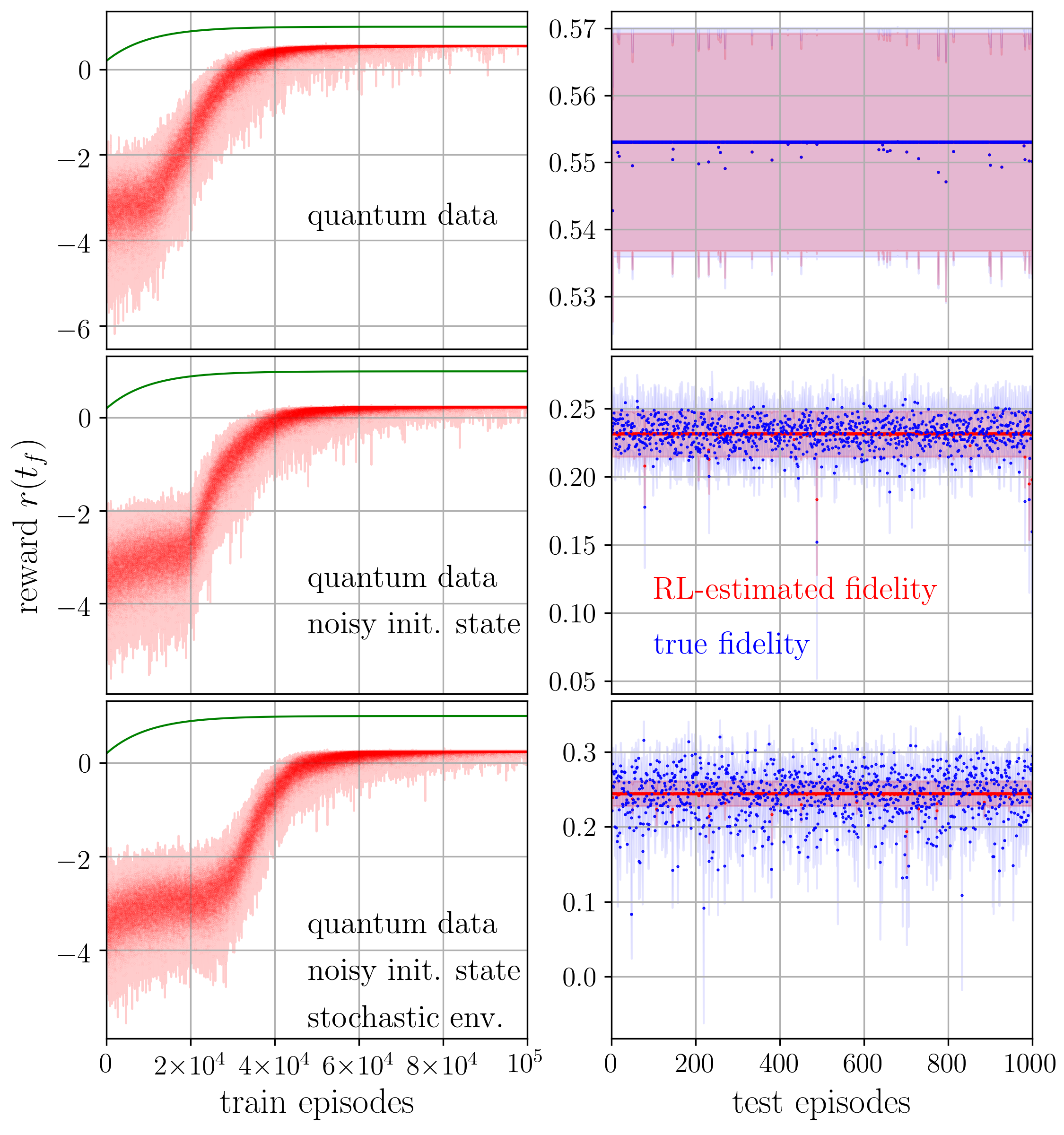}
	\caption{\label{fig:fid_traces_classical}Training behavior as a function of the number of Train episodes for the classical Kapitza pendulum. The oscillator parameters are $\Omega/\omega_0=10$, $A=2$ and $m\omega_0=1$. The protocols contain $8$ steps per period for a total of $N_T=4$ periods. The target is $\theta_\ast=\pi$ and $p_\ast=0$. See \href{https://mgbukov.github.io/movies/RL_kapitza/movie-3.mp4}{Video 3} for a visualization of the best-encountered RL protocol.}	
\end{figure}

\section{\label{app:algo}Q-Learning Algorithm for Autonomous Quantum Control}

Below, I provide a trivial extension of Watkins tabular Q-Learning~\cite{sutton1998reinforcement} which uses as a reward the noisy running estimate of the fidelity $F_h(t_f)$, and learns from experience replays.

To study the Floquet control problem, I apply the version of a tabular Q-Learning algorithm~\cite{sutton1998reinforcement} with eligibility trace depth parameter $\lambda=0.6$. For the Q-Learning update rule, I use a small learning rate of $\alpha=0.1$ in order to account for the running stochastic reward (the fidelity), estimated from binary quantum measurements as $r=m/n$. Here $m$ is the number of $+1$ measurement outcomes, and $n$ is the total number of measurements for a fixed protocol [see main text]. To gain measurement statistics, each protocol is repeated $100$ times every time it is encountered, until the error estimate to be within the $2\sigma$-window, $E=2\sqrt{r(1-r)/n}$, becomes less than $1\%$. During this repetition stage, no updates of the $Q$-function take place. To choose actions, the algorithm uses an $\varepsilon$-greedy policy: the best action [according to the current $Q$-function] is taken with probability $1-\varepsilon(n_\mathrm{ep})$, or else a random action is chosen with probability $\varepsilon(n_\mathrm{ep})$. The exploration schedule $\varepsilon(n_\mathrm{ep})$ is attenuated exponentially according to
\begin{equation}
\varepsilon(n_\mathrm{ep})=(\varepsilon_i-\varepsilon_f)\; \exp\left(-\frac{10n_\mathrm{ep}}{N_\mathrm{episodes}}\right) + \varepsilon_f,
\end{equation}
with $\varepsilon_i=10$ and $\varepsilon_f=50$ (chosen empirically). The larger $\varepsilon(n_\mathrm{ep})$, the less the RL agent explores [see green curves in all learning plots, where $\varepsilon(n_\mathrm{ep})$ is normalized within $[0,1]$ for display purposes]. The current episode and the total number of train episodes are denoted by $n_\mathrm{ep}$ and $N_\mathrm{episodes}$, respectively. In order to help the agent explore the exponentially large RL state space efficiently, I keep track of the best encountered protocol w.r.t.~the current fidelity estimate, and replay it every $100$ episodes for $200$ times, thereby updating the $Q$-function. 

Algorithm~\ref{algo:QQL}, describes the pseudo-code for the RL algorithm used to obtain the results in the main text. Familiarity with the original Watkins' Q-Learning algorithm and its extension TD$(\lambda)$, see e.g.~Ref.~\cite{sutton1998reinforcement}, is helpful to facilitate understanding. It is straightforward to extend Algorithm~\ref{algo:QQL} to Deep Learning.

\begin{algorithm}
	\caption{Q-Learning with nondeterministic rewards [quantum measurements]}\label{algo:QQL}
	\begin{algorithmic}[1]
		\Procedure{Q-Learning}{}
		\State initialize an empty $Q(s, a)$ function for all states $s\in\mathcal{S}$ and actions $a\in\mathcal{A}(s)$
		\State initialize an empty registry $R(h)=(m,n,E)$ for all protocol sequences $h$, number of protocol encounters $m\in\mathbb{N}$, number of positive quantum measurement outcomes $n\in\mathbb{N}$, and statistical error estimate $E\in\mathbb{R}$
		\State initialize \texttt{best\_encountered\_actions} arbitrarily and find the corresponding \texttt{best\_encountered\_protocol}
		\State initialize \texttt{best\_encountered\_return}$=-1$
		\Repeat { for every episode:}
		\State set value of $\varepsilon$-greedy exploration according to some schedule
		\State run $\varepsilon$-greedy \texttt{QL\_episode} in \texttt{explore} mode
		\If {statistical error estimate $E$ of return for most recent protocol $h$ is within some threshold}
		\State repeat protocol to collect data and improve statistics (Q-function is not updated)
		\State update registry $R(h)$.
		\EndIf
		\State run \texttt{QL\_episode} in \texttt{repeat} mode (Q-function is updated)
		\State run \texttt{update\_best\_encountered} routine
		
		\If {episode is scheduled for replay}
		\If{statistical error estimate $E$ of return for best protocol $h$ is within some threshold}
		\Repeat 
		\State run greedy \texttt{QL\_episode} in \texttt{replay} mode with unit learning rate $\alpha=1$ (Q-function is updated)
		\State \texttt{best\_encountered\_return} $\gets r$
		\Until {a number of replay episodes is exhausted}
		\EndIf
		\EndIf
		
		\Until{number of episodes is exhausted}\\ \\
		\EndProcedure
	\end{algorithmic}
	
	\begin{algorithmic}[1]
		\Function{\texttt{QL\_episode(mode)}}{}
		\State reset environment into initial state $S=S_0$
		\State $q\gets Q(S_0,:)$ (compute Q-function value of initial state for all available actions $a\in\mathcal{A}(s)$)
		\State $\mathrm{trace}(s,a) \equiv 0, \ \mathrm{for\ all}\ s\in\mathcal{S}, a\in\mathcal{A}(s)$
		
		\Repeat{ for each step in the episode:}
		\State run \texttt{choose\_action(mode)} to get action $A$ from state $S$ using policy derived from $Q$ (e.g., $\varepsilon$-greedy)
		\State take action $A$, environment goes to new state $S'$
		\State $\mathrm{trace}(S,A) \gets \alpha$: fire eligibility trace
		\State set $\delta_t \gets - q(A)$
		
		\If {$S$ is terminal}
		\State compute current estimate of return $r$
		\State $\delta_t\gets \delta_t + r$
		\State $Q(s,a)\gets Q(s,a) + \delta_t\;\mathrm{trace}(s,a)$ for all $s\in\mathcal{S}, a\in\mathcal{A}(s)$
		\State \textbf{goto} next episode
		\EndIf
		
		\State $q\gets Q(S',:)$ (find Q-function for all actions in state $S'$)
		\State $\delta_t \gets \delta_t + \max_a q(a)$ (find action that maximizes $q$)
		\State $Q(s,a)\gets Q(s,a) + \delta_t\;\mathrm{trace}(s,a)$ for all $s\in\mathcal{S}, a\in\mathcal{A}(s)$
		\State $\mathrm{trace}(s,a)\gets \lambda\; \mathrm{trace}(s,a)$ for all $s\in\mathcal{S}, a\in\mathcal{A}(s)$
		
		\Until{episode is complete}\\ \\
		\EndFunction
	\end{algorithmic}

	\begin{algorithmic}[1]
		\Function{\texttt{choose\_action(mode)}}{}
		
		\If{\texttt{mode} is \texttt{repeat}}
		\State $A\gets$ action taken in previous run at this time step
		\ElsIf{\texttt{mode} is \texttt{explore}}
		\State compute available actions $\mathcal{A}(S)$ from current state $S$
		\State compute $A_\mathrm{greedy}\gets $ one of the (possibly many) actions that maximize $Q(S,:)$
		\State choose $A\gets A_\mathrm{greedy}$ with probability $\varepsilon$ (depending on exploration schedule), otherwise $A\gets$ a random available action
		
		\If {$A$ is not $A_\mathrm{greedy}$}
		\State $\mathrm{trace}(s,a)\gets 0$, reset trace for all $s\in\mathcal{S}, a\in\mathcal{A}(s)$
		
		\EndIf
		\ElsIf{\texttt{mode} is \texttt{replay}}
		\State $A\gets$ \texttt{best\_encountered} action at this time step			
		\EndIf
		
		\EndFunction
	\end{algorithmic}
	
	\begin{algorithmic}[1]
		\Function{\texttt{update\_best\_encountered}}{}
		\If {\texttt{return\_best\_encountered}$<$current return $r$}
		\If {statistical error estimate $E$ of return for most recent protocol $h$ is within some threshold}
		\State \texttt{return\_best\_encountered} $\gets r$
		\State overwrite \texttt{actions\_best\_encountered} with most recent actions sequence
		\State overwrite \texttt{state\_best\_encountered} with most recent protocol
		\EndIf	
		\EndIf	
		\EndFunction
	\end{algorithmic}
	
\end{algorithm}

\section{\label{app:videos}Video Simulations of the RL-Controlled Kapitza Oscillator}

Legends for all three movies is available on \href{https://mgbukov.github.io/RL\_kapitza/}{https://mgbukov.github.io/RL\_kapitza/}.

\end{widetext}
\end{document}